\documentclass[12pt]{article}
\usepackage{epsf}

\newcommand{\beq}{\begin{equation}}
\newcommand{\eeq}{\end{equation}}
\newcommand{\ba}{\begin{array}}
\newcommand{\ea}{\end{array}} 
\newcommand{\beqa}{\begin{eqnarray}}
\newcommand{\eeqa}{\end{eqnarray}}
\newcommand{\dis}{\displaystyle}
\newcommand{\cL}{{\cal L}}

\newcommand{\cA}{{\cal A}}
\newcommand{\cO}{{\cal O}}
\newcommand{\cM}{{\cal M}}
\newcommand{\Br}{{\cal B}}
\newcommand{\da}{^\dagger}
\newcommand{\no}{\nonumber}
\newcommand{\lsim}{\stackrel{<}{_\sim}}
\newcommand{\gsim}{\stackrel{>}{_\sim}}
\newcommand{\eps}{\varepsilon}

\newcommand{\tu}{{\tilde u}}

\newcommand{\ttop}{{\tilde t}}

\newcommand{\op}{Q}
\newcommand{\pk}{p_{_K}}
\newcommand{\Ks}{{K^*}}
\newcommand{\cne}{C_9^{\rm eff}}
\renewcommand{\Im}{{\rm Im}}
\renewcommand{\Re}{{\rm Re}}
\newcommand{\ct}{C_{10}}

\def\npb#1#2#3{    {\it Nucl. Phys. }{\bf B #1} (#2) #3}
\def\plb#1#2#3{    {\it Phys. Lett. }{\bf B #1} (#2) #3}
\def\prd#1#2#3{    {\it Phys. Rev. }{\bf D #1} (#2) #3}
\def\prep#1#2#3{   {\it Phys. Rep. }{\bf #1} (#2) #3}
\def\prl#1#2#3{    {\it Phys. Rev. Lett. }{\bf #1} (#2) #3}

\def\rmp#1#2#3{    {\it Rev. Mod. Phys. }{\bf #1} (#2) #3}
\def\zpc#1#2#3{    {\it Zeit. f{\"u}r Physik }{\bf C #1} (#2) #3}
\def\mpla#1#2#3{   {\it Mod. Phys. Lett. }{\bf A #1} (#2) #3}

\def\epjc#1#2#3{   {\it Eur. Phys. J. }{\bf C #1} (#2) #3}
\def\ibid#1#2#3{   {\it ibid. }{\bf #1} (#2) #3}
\def\jhep#1#2#3{   {\it JHEP  }{\bf #1} (#2) #3} 

\def\o{{\cal O}}
\def\c{{C}}

\begin{document}

\thispagestyle{empty}
\setcounter{page}{0}
\begin{flushright}
CERN-TH/2000-156\\
SLAC-PUB-8430\\
TUM-HEP-374/00\\
May 2000
\end{flushright}
\vspace*{1.5cm}
\centerline{\Large\bf Phenomenology of non-standard $Z$ couplings}
\vspace*{0.3cm}
\centerline{\Large\bf in exclusive semileptonic $b\to s$ 
 transitions\footnote{Work supported by the Department of Energy, 
 Contract DE-AC03-76SF00515} }
\vspace*{2cm}
\centerline{{\sc Gerhard Buchalla${}^a$}, {\sc Gudrun Hiller${}^b$} and  
            {\sc Gino Isidori${}^{c,d}$}}
\bigskip
\centerline{\sl ${}^a$Theory Division, CERN, CH-1211 Geneva 23,
                Switzerland}
\centerline{\sl ${}^b$Stanford Linear Accelerator Center, 
              Stanford University, Stanford, CA 94309, USA }
\centerline{\sl ${}^c$ Physik Department, Technische Universit{\"a}t 
              M{\"u}nchen, D-85748 Garching, Germany}
\centerline{\sl ${}^d$INFN, Laboratori Nazionali di Frascati, 
                I-00044 Frascati, Italy}
\vspace*{1.5cm}
\centerline{\bf Abstract}
\vspace*{0.3cm}
\noindent 
The rare decays $B\to K^{(*)}\ell^+\ell^-$,
$B\to K^{(*)}\nu\bar\nu$ and $B_s\to\mu^+\mu^-$
are analyzed in 
a generic scenario where New Physics effects enter predominantly
via $Z$ penguin contributions. We show that  
this possibility is well motivated
on theoretical grounds, as the $\bar sbZ$ vertex is particularly
susceptible to non-standard dynamics. In addition,
such a framework is also interesting phenomenologically since
the $\bar sbZ$ coupling is rather poorly constrained by present data.
The characteristic features of this scenario
for the relevant decay rates and distributions
are investigated.
We emphasize that both sign and magnitude of the
forward-backward asymmetry of the decay leptons in
$\bar B\to \bar K^*\ell^+\ell^-$, ${\cal A}^{(\bar B)}_{FB}$, 
carry sensitive information on New Physics. The observable
${\cal A}^{(\bar B)}_{FB}+{\cal A}^{(B)}_{FB}$ is proposed as
a useful probe of non-standard CP violation in $\bar sbZ$ couplings.

\vfill

\newpage
\pagenumbering{arabic}

\section{Introduction}
Despite the fact that the Cabibbo-Kobayashi-Maskawa (CKM) mechanism
provides a consistent description of presently 
available data on quark-flavour mixing, the flavour 
structure of the Standard Model (SM) is not very 
satisfactory from the theoretical point of view,
especially if compared to the elegant and economical 
gauge sector. On the contrary, it is natural to consider 
it as a phenomenological low-energy description of a more 
fundamental theory, able, for instance, to explain 
the observed hierarchy of the CKM matrix. 

A special role in searching for experimental clues about 
non-standard flavour dynamics is provided by flavour-changing 
neutral-current (FCNC) processes. Within the SM these 
are generated only at the quantum level and are additionally 
suppressed by the smallness of the off-diagonal entries of 
the CKM matrix. On one side this makes their observation very
challenging but on the other side it ensures a large sensitivity 
to possible non-standard effects, even if these occur at 
very high energy scales.

In general we can distinguish two types of FCNC processes:
$\Delta F=2$ and $\Delta F=1$ transitions. The former has 
been successfully tested in $K^0-\bar{K}^0$ and 
$B_d-\bar{B_d}$ systems, both via $CP$-conserving 
($\Delta M_K$ and $\Delta M_{B_d}$) and $CP$-violating 
observables ($\varepsilon_K$ and $\sin 2\beta$). 
On the other hand, much less is known about the latter.
Few $\Delta S=1$ FCNC transitions have been observed 
in $K$ decays, but most of them are affected 
by sizable long-distance uncertainties.
The only exception is $\Br(K^+\to \pi^+ \nu\bar{\nu})$ \cite{E787}, 
which is however affected by a large experimental error.
The situation is slightly better in the $B$ sector, 
where the inclusive $b \to s\gamma$ rate
provides a theoretically clean
$\Delta B=1$ FCNC observable \cite{bsgamma}.
Nonetheless, it is clear that a substantial 
improvement is necessary in order to perform 
more stringent tests of the SM. 

In the present paper we focus on a specific class of non-standard
$\Delta B=1$ FCNC transitions: those mediated by
the $Z$-boson exchange. As we shall discuss, these 
are particularly interesting for two main reasons:
i) there are no stringent experimental bounds 
   on these transitions yet;
ii) it is quite natural to conceive extensions of 
the SM where the $Z$-mediated FCNC amplitudes
are substantially modified, even taking into 
account the present constraints on $\Delta B=2$
and $b \to s\gamma$ processes.

The simplest way to search for non-standard  $\Delta B=1$
FCNC effects mediated by the $Z$-boson exchange
is to look for parton-level transitions of the type
$b \to s(d) + \ell^+\ell^-(\nu\bar\nu)$. 
None of such processes has been observed yet, but 
the situation will certainly improve in a short term, 
with the advent of new high statistics experiments
at $e^+e^-$ and hadron $B$-factories. 
In principle the theoretically cleanest observables are
provided by inclusive decays, which should play an important
role in the longer run.
On the other hand, the exclusive variants will be more readily
accessible in experiment. Despite the sizable theoretical
uncertainties in the exclusive hadronic form factors, these
processes could therefore give interesting first clues on deviations
from what is expected in the Standard Model. This is particularly true
if those happen to be large or if they show striking patterns.
Since in the present study we are mainly interested in such a
possibility,
we shall restrict our phenomenological 
discussion to the exclusive three-body processes
$B \to (K,K^*) + (\mu^+\mu^-, \nu\bar\nu)$. Having 
branching ratios in the $10^{-6}-10^{-5}$ range,
and a relatively clear signature, these decays
represent one of the primary goals of the 
new experiments. As we will show, forward-backward 
and $CP$ asymmetries of these modes provide a 
powerful tool not only to search for New Physics, 
but also to clearly identify the interesting 
scenario where the dominant source of non-standard 
dynamics can be encoded in effective FCNC 
couplings of the $Z$ boson.

The paper is organized as follows. 
In Section 2 the general features 
characterizing the FCNC couplings of the $Z$ boson
beyond the SM are discussed; we further introduce a general 
parameterization 
of these effects, both for $b \to s$ and $b \to d$ transitions,
in terms of the complex couplings $Z^{L,R}_{qb}$ ($q=s,~d$)
and evaluate their model-independent constraints.
In Section 3 we present various estimates for these 
couplings in specific extensions of the Standard Model.
Notations and general formulae for the 
phenomenological analysis are introduced in Section 4. 
In Section 5 and Section 6 we discuss how the non-standard 
FCNC couplings of the $Z$ would manifest themselves
and how they could possibly be isolated in 
$B \to (K,K^*) + \nu\bar\nu$ and $B \to (K,K^*) + \mu^+\mu^-$
decays, respectively. Implications for $B_s\to\mu^+\mu^-$
are briefly described in Section 7. 
A summary of the results can be found in Section 8.

\section{General features of FCNC couplings of the $Z$ boson}

In a generic extension of the Standard Model where new 
particles appear only above some high scale $M_X > M_Z$, 
we can always integrate out the new degrees of freedom and generate 
a series of local FCNC operators already at the electroweak scale.
Those relevant for $b \to s(d) + \ell^+\ell^- (\nu\bar{\nu})$
transitions can be divided into three wide classes:

\begin{itemize}
\item{} {\it Four-fermion operators}. 
The local four-fermion operators obtained by integrating 
out the new particles necessarily have dimension greater 
or equal to six. 
These could be generated  either at the tree level 
(e.g. by leptoquark exchange) or at one loop (e.g. by SUSY 
box diagrams) but in both cases, due to dimensional arguments,
their Wilson coefficients are expected to be suppressed at 
least by two inverse powers of the New Physics scale $M_X$. 

\item{} {\it Magnetic operators}. 
The integration of the heavy degrees of freedom 
can also lead to operators with dimension lower 
that six, creating an effective FCNC coupling 
between quarks and SM gauge fields. 
In the case of the photon field, 
the unbroken electromagnetic gauge 
invariance implies that the lowest dimensional coupling 
is provided by the so-called ``magnetic'' operators
$\sim \bar{b}\sigma^{\mu\nu} s F_{\mu\nu}$.
Having dimension five, their Wilson coefficients 
are expected to be suppressed at least 
by one inverse power of $M_X$.

\item{} {\it FCNC $Z$ couplings}. 
Due to the spontaneous breaking of $SU(2)_L \times U(1)_Y$
we are allowed, in the case of the $Z$ boson, to build an 
effective FCNC coupling of dimension four: 
$\bar{b}_{L(R)} \gamma^\mu s_{L(R)} Z_\mu$.
The coefficient of this operator must be 
proportional to some symmetry-breaking 
term but, for dimensional reasons, it does not need 
to contain any explicit $1/M_X$ suppression.
\end{itemize}

\noindent
Given the above discussion, the effective FCNC 
couplings of the $Z$ boson appear particularly interesting 
and worth to be studied independently of the other effects: 
in a generic model with additional 
sources of $SU(2)_L \times U(1)_Y$ breaking, 
these are the only $\Delta F=1$ FCNC couplings that 
do not necessarily decouple by dimensional arguments
in the limit $M_X/M_Z \gg 1$. 
It should be noticed 
that the requirement of naturalness in the size of the 
$SU(2)_L \times U(1)_Y$ breaking terms suggests that also 
the adimensional couplings of the 
non-standard $Z$-mediated FCNC amplitudes
decouple in the limit $M_X/M_Z \to \infty$. However, the above naive 
dimensional argument remains a strong indication 
of an independent behaviour of these couplings with respect 
to the other FCNC amplitudes \cite{CI,SIL}.
As we will illustrate in Section~\ref{sect:3}, this 
independent behaviour is indeed realized within various 
extensions of the SM.

Interestingly, FCNC couplings of the $Z$ represent 
also the least constrained class among those listed
above: magnetic operators are bounded 
by $b \to s\gamma$ and, within most models,
dimension-six operators are strongly correlated 
to those entering $B -\bar B$ mixing.
The scenario where the dominant non-standard 
contribution to $b \to s(d) + \ell^+\ell^- (\nu\bar{\nu})$
transitions is mediated by a $Z\bar{b}s(d)$ coupling
is therefore particularly appealing also 
from a purely phenomenological point of view.

\subsection{Effective Lagrangian and model-independent constraints}
The effective FCNC couplings of the $Z$, relevant for the 
$b\to s$ transition, can be described by means 
of the following effective Lagrangian
\beq
  \label{eq:Zsb}
  \cL^{Z}_{FC} = \frac{G_F}{\sqrt{2}} \frac{e}{ \pi^2} M_Z^2
  \frac{\cos \Theta_W}{\sin \Theta_W} Z^\mu \left( 
   Z^L_{sb}~\bar b_L \gamma_\mu s_L + 
   Z^R_{sb}~\bar b_R \gamma_\mu s_R \right) \,+\, {\rm h.c.}~,
\eeq
where $Z^{L,R}_{sb}$ are complex couplings and the overall
normalization has been chosen in analogy to the $s\to d$ 
case discussed in \cite{CI,BS}. 
For later convenience we also 
define $Z^{L,R}_{bs}=(Z^{L,R}_{sb})^*$.
The SM contribution 
to $Z^{L,R}_{sb}$, evaluated in the  't~Hooft-Feynman gauge, 
can be written as\footnote{~As it is well known, the 
SM contribution to FCNC $Z$ penguins is not
gauge invariant. We recall, however, that the leading 
contribution to both $b \to s(d) \ell^+\ell^- $ and  
$b \to s(d) \nu\bar\nu $ amplitudes in the limit $x_t \to \infty$ 
is gauge independent and is indeed generated by the $Z$ penguin
($C_0(x_t) \to x_t/8$ for $x_t\to\infty$).}
\beq
  \label{eq:SMZsb} 
  Z^R_{sb}\vert_{\rm SM} = 0~,\qquad
  Z^L_{sb}\vert_{\rm SM} = V_{tb}^* V_{ts} C_0(x_t)~,
\eeq
where $V_{ij}$ denote the CKM matrix elements,
$x_t=m_t^2/m_W^2$ and the function $C_0(x)$ can be 
found in \cite{BBL}. 

At present the cleanest model-independent constraints on 
$|Z^{L,R}_{sb}|$ can be obtained from the experimental 
upper bounds on $\Br(B\to X_s \ell^+\ell^-)$. 
Normalizing the inclusive rate for $B\to X_s \ell^+\ell^-$ to the 
well known  $\Gamma(B\to X_c e^+ \nu_e)$ and assuming that all
contributions to the former but those generated by  ${\cal L}^{Z}_{FC}$
are negligible, we can write
\beq
\dis\frac{\Gamma(B\to X_s \ell^+\ell^-)}{\Gamma(B\to X_c  e^+ \nu_e)}
= \dis\frac{\alpha^2}{\pi^2 \sin^4\Theta_W}~
  \frac{\left| Z^L_{sb} \right|^2 + 
  \left| Z^R_{sb} \right|^2}{\left|V_{cb}\right|^2 f(m_c/m_b)}
  \left[ \left( a_L^\ell \right)^2 +\left( a_R^\ell \right)^2 \right]~,
  \label{eq:inclrate}
\eeq
where  $f(z)=(1-8z^2+8z^6-z^8-24z^4\ln z)$ 
is the phase space factor due to the non-vanishing charm mass 
and, for consistency, we have neglected the small QCD correction factor
in $\Gamma(B\to X_c  e^+ \nu_e)$. Here
$a^\ell_{L(R)}$ denotes the left(right)-handed coupling of the lepton 
to the $Z$, namely $a^\ell_L=\sin^2\Theta_W-1/2$ and 
$a^\ell_R=\sin^2\Theta_W$ for $\ell=e$ or $\mu$, whereas
$a^\nu_L=1/2$ and $a^\nu_R=0$ for the neutrino case.
Using $\Br(B\to X_c  e^+ \nu_e)=0.105$, $\sin^2\Theta_W=0.23$, 
$\alpha^{-1}=129$,
$\left|V_{cb}\right|=0.04$ and $f(m_c/m_b)=0.54$, we find 
\beqa
\Br(B\to X_s \ell^+ \ell^- ) &=& 1.76 \times 10^{-3}~ \left(  
\left| Z^L_{sb} \right|^2 + \left| Z^R_{sb} \right|^2 \right)~, \\
\Br(B\to X_s \nu \bar{\nu} ) &=& 1.05 \times 10^{-2}~ \left(  
\left| Z^L_{sb} \right|^2 + \left| Z^R_{sb} \right|^2 \right)~,
\eeqa
where in the neutrino mode we have summed over the three lepton families.
Experimental upper bounds exist both for $\Br(B\to X_s\ell^+\ell^-)$ 
and $\Br(B\to X_s \nu \bar{\nu} )$, leading to 
\beqa
 \left( \left| Z^L_{sb} \right|^2 + \left| Z^R_{sb} \right|^2 \right)^{1/2}
   \lsim 0.15~, && \qquad {\rm from}\qquad\quad
    \Br(B\to X_s \ell^+ \ell^- ) < 4.2 \times 10^{-5}\
    \protect\cite{CLEO}~, \label{eq:Zsblim} \\
 \left( \left| Z^L_{sb} \right|^2 + \left| Z^R_{sb} \right|^2 \right)^{1/2}
   \lsim 0.27~,  && \qquad {\rm from}\qquad\quad
    \Br(B\to X_s \nu \bar{\nu} ) < 7.7 \times 10^{-4}\
     \protect\cite{ALEPH}~.  \label{eq:Zsblim2}
\eeqa
The strongest bound is presently imposed by 
$\Br(B\to X_s\ell^+\ell^-)$,
since the larger sensitivity of $\Br(B\to X_s \nu \bar{\nu} )$  
is compensated by its more difficult experimental 
determination.\footnote{~A result similar to the one in 
(\ref{eq:Zsblim}) has recently been presented also in Ref.~\cite{GKN}.}
The limits in (\ref{eq:Zsblim}--\ref{eq:Zsblim2}) have been derived assuming 
that all the non-$Z$-mediated contributions  are 
negligible, which is a reasonable approximation in view 
of the present experimental sensitivities. On the other hand, 
if the experimental bounds were much closer to the SM expectations, 
we stress that the neutrino mode would definitely be preferable from the 
theoretical point of view due to the absence of 
electromagnetic and long-distance contributions \cite{GLN,BIR}.

Employing the Wolfenstein expansion of the 
CKM matrix in powers of $\lambda = 0.22$ \cite{wolfCKM}
and recalling that $C_0(x_t) \sim \cO(1)$, 
the SM contribution to $Z^L_{sb}$ turns out to be 
of $\cO(\lambda^2)\sim 0.04$ (see Eq.~(\ref{eq:SMZsb})), 
therefore much below the bound (\ref{eq:Zsblim}).
As we will show later, 
more severe constraints on  $| Z^{L,R}_{sb} |$
can be obtained by the experimental bound on the exclusive 
branching ratio $\Br(B\to K^* \mu^+\mu^-)$. These are however 
subject to stronger theoretical uncertainties, related to 
the assumptions on the form factors, and require 
a detailed discussion that we postpone
to Section~\ref{sect:BRKpll}.

Additional model-independent information on these couplings 
could in principle be obtained by the direct constraints 
on $\Br(Z\to b \bar s)$ and by  $B_s- \bar B_s$ mixing,
but in both cases these are not very significant.
Concerning the first case, we find 
\beq
\Br(Z\to b \bar s) ~=~ \frac{G_F^2 M_Z^5 \alpha \cos^2\theta_W}{4 \pi^4 
\Gamma_Z \sin^2\theta_W} \left(  
\left| Z^L_{sb} \right|^2 + \left| Z^R_{sb} \right|^2 \right)~=~
2.3 \times 10^{-5}~ \left(  
\left| Z^L_{sb} \right|^2 + \left| Z^R_{sb} \right|^2 \right)~,
\label{eq:Ztobs}
\eeq
which is quite far from the present experimental sensitivity
at LEP of ${\cal O}(10^{-3})$ \cite{L3}, 
even for $| Z^{L,R}_{sb}|\sim \cO(1)$. 
Using the bound (\ref{eq:Zsblim}) in Eq.~(\ref{eq:Ztobs}) leads to
\beq
\Br(Z\to b \bar s) ~\lsim~ 5 \times 10^{-7}~,
\eeq
to be compared with the SM expectation 
$\Br(Z\to b \bar s)\vert_{\rm SM} \simeq 1.4 \times 10^{-8}$ 
\cite{Ztobs}.  

Concerning $B_s- \bar B_s$ mixing, assuming for 
simplicity $Z^R_{sb}=0$ and employing the notations of \cite{BBL}, we find
\beqa
\cM(B_s - \bar B_s)^Z &=& \frac{\alpha G_F^2 M_W^2}{3 \pi^3 \sin^2\theta_W}
  B_{B_s} f^2_{B_s} M_{B_s} \eta_B \left(Z^L_{sb}\right)^2 \\
 &=& \frac{ 4 \alpha \cM(B_s - \bar B_s)^{SM} }{\pi \sin^2\theta_W S_0(x_t) }
     \left(\frac{Z^L_{sb}}{V_{tb}^*V_{ts}}\right)^2~.
\label{eq:BBmix}
\eeqa
At the moment  we cannot extract any interesting information from 
(\ref{eq:BBmix}) due to the lack of a 
significant upper bound on $|\cM(B_s - \bar B_s)|$.
If in the future we were able to exclude that 
$|\cM(B_s - \bar B_s)^Z|$ is larger than 
$|\cM(B_s - \bar B_s)^{SM}|$, then we would obtain
\beq
 \label{eq:ZsblimBB}
  \left| Z^L_{sb} \right| < 7.6  \left| V_{tb}^*V_{ts} \right|
  \sim 0.3~.
\eeq

Performing the exchange $s\to d$ in Eq.~(\ref{eq:Zsb}-\ref{eq:SMZsb}) 
we can define, analogously to $Z^{L,R}_{sb}$,
the couplings $Z^{L,R}_{db}$ relevant for the $b\to d$ transition. 
The upper bound (\ref{eq:Zsblim}) would be valid also 
for these couplings if we could 
assume $\Br(B\to X_d \mu^+\mu^-)\leq \Br(B\to X_s \mu^+\mu^-)$, 
but in the $b\to d$ case more stringent constraints
can be derived from $B_d - \bar B_d$ mixing. 
The SM contribution to $\cM(B_d - \bar B_d)$ can 
account for the observed value of $\Delta M_{B_d}$,
nevertheless, due to the theoretical uncertainty on
$B_{B_d} f^2_{B_d}$, non-standard contributions of 
comparable size cannot be excluded at present.
Imposing for instance $|\cM(B_d - \bar B_d)^Z| <
|\cM(B_d - \bar B_d)^{SM}|$ and replacing 
$s\to d$ in Eq.~(\ref{eq:BBmix}) we obtain
\beq
 \label{eq:Zsdlim}
  \left| Z^L_{db} \right| < 7.6  \left| V_{tb}^*V_{td} \right|
  \sim 0.06~,
\eeq
which is still substantially larger than the SM 
contribution: $Z^L_{db}\vert_{\rm SM} =\cO(\lambda^3)\sim 0.01$.

\section{Model-dependent expectations for $Z^{L,R}_{qb}$ }
\label{sect:3}

In the previous section we have seen that 
sizable non-standard contributions to the 
FCNC couplings of the $Z$ are allowed, 
at least from a purely phenomenological point of view,
both for $b \to s$ and $b \to d$ transitions.
In the following we shall analyze the expectations 
for the $Z^{L,R}_{qb}$ couplings in a few
specific theoretical frameworks. Moreover, we will show 
various consistent models where it is a good approximation to 
encode all the non-standard FCNC effects in the 
couplings of ${\cal L}^{Z}_{FC}$ .

\subsection{Fourth generation}
A simple extension of the SM, particularly 
useful as a toy model for more complicated scenarios,
is obtained by adding a sequential fourth 
generation of quarks and leptons.
This is allowed by Tevatron and LEP data provided all the new 
fermions, neutrinos included, 
are sufficiently heavy ($m_{t'} \gsim 200$ GeV) 
and the splitting among the weak isospin 
doublets is very small ($|m_{t'}-m_{b'}|/m_{t'} \lsim 0.1$)
(see e.g. \cite{PQ} and references therein).

This model exhibits a typical
non-decoupling effect in the $Z_{qb}$ coupling.
Indeed, denoting by $V_{t'q}$ the mixing angles of the new
up-type quark with the light generations, 
the dominant non-standard contribution 
to the $Z^{L,R}_{qb}$ coupling is given by 
\beq
  \label{eq:4thZsb} 
  Z^R_{qb}\vert_{4^{\rm th}} = 0~,\qquad
  Z^L_{qb}\vert_{4^{\rm th}} = 
 V_{t'b}^* V_{t'q} C_0(x_{t'}) \simeq  \frac{x_{t'}}{8} V_{t'b}^* V_{t'q}~,
\eeq
where $x_{t'}=m_{t'}^2/m_W^2$. 
In the limit 
\beq
V_{t'b}^* V_{t'q} \to 0~, \qquad  
m^2_{t'} \to \infty~, \qquad  
V_{t'b}^* V_{t'q} m^2_{t'} \to {\rm const.}, 
\eeq
this is the only non-standard effect 
surviving in $b \to s(d)  + \ell^+\ell^- (\nu\bar{\nu})$
transitions.
Choosing sufficiently small mixing angles
one can therefore easily evade the experimental 
constraint on $V_{t'b}^* V_{t'q}$ and, by raising the 
value of $m_{t'}$, still obtain sizable effects 
in $Z^{L}_{qb}$.

In the case of $b \to s$ transitions
the dominant constraint on the combination 
$V_{t'b}^* V_{t's}$ is imposed by $b \to s \gamma$. 
Indeed the bounds from $K-\bar K$ mixing and $K$ decays 
can always be evaded assuming $V_{t'd}=0$, 
whereas the constraint from $B_s- \bar B_s$ mixing
is very loose. Barring accidental cancellations 
in the $b \to s\gamma$ amplitude, namely assuming that the 
dominant contribution to the latter is the SM one,
leads to $\left|V_{t'b}^* V_{t's}\right| \lsim \lambda^3$, 
almost independently of the value of $m_t'$.
Even employing this stringent constraint,\footnote{~Substantially larger 
values of $\left|V_{t'b}^* V_{t's}\right|$ are possible assuming that 
the contribution of the 
fourth generation changes the sign of the $b \to s\gamma$ amplitude. 
See Ref.~\protect\cite{fourth}
for a recent discussion of this point.}
however, one could still have $|Z^L_{sb}\vert_{4^{\rm th}}| \sim 
|Z^L_{sb}\vert_{\rm SM}|$ provided $m_{t'}\gsim 400$~GeV.

\subsection{Generic SUSY models}
\label{sect:susy}

Due to the large number of new particles carrying flavor quantum 
numbers, sizable modifications of FCNC amplitudes
are naturally expected within low-energy supersymmetric 
extensions of the SM with generic flavour couplings \cite{SusyFCNC,HKR}.
Assuming $R$ parity conservation and  
minimal particle content, FCNC amplitudes 
involving external quark fields turn out to 
be generated only at the quantum level, like in the SM. 
However, in addition to the standard penguin and box dia\-grams, also their
corresponding superpartners, generated by gaugino/higgsino-squark 
loops, play an important role. These contributions to 
inclusive and exclusive $b\to s \ell^+\ell^-$ transitions have been widely 
discussed in the literature (see e.g. \cite{LMSS,ABHH99,LS,KSsusy,Qishu} 
for a recent discussion and a complete list of references),
employing different assumptions for the soft-breaking terms.
In the following we will emphasize the role of the $Z$
penguins in the context of the mass-insertion 
approximation \cite{HKR}.

Similarly to the $Z\bar{s}d$ case, extensively discussed 
in \cite{CI,MWBRS}, the potentially dominant non-SM effect
in the effective $Z\bar{b}q$ vertex is generated by  
chargino-up-squark diagrams \cite{LMSS,LS}. 
Indeed sizable $SU(2)_L$ breaking effects
can be expected only in the up sector due 
to the large Yukawa coupling of the third generation.
Moreover, since terms involving external right-handed quarks are 
suppressed by the corresponding down-type Yukawa couplings,
also within this framework $Z^{R}_{qb}$
turns out to be negligible. 

Employing the notations of \cite{CI}, the full
chargino-up-squark contribution to $Z^{L}_{sb}$
can be written as 
\beq
Z_{sb}^{L}\vert_{\rm SUSY} = {1\over 8} A^s_{jl}{\bar A}^b_{ik}
F_{jilk}~,
\label{eq:Zchi}
\eeq
where 
\beqa
A^s_{jl} &=& {\hat H}_{l s_L}{\hat V}\da_{1j} 
            - g_t V_{ts}{\hat H}_{l t_R}{\hat V}\da_{2j}~, \\
{\bar A}^b_{ik} &=& {\hat H}\da_{b_L k}{\hat V}_{i1} 
            - g_t V_{tb}^*{\hat H}\da_{t_R k}{\hat V}_{i2}~, \\
F_{jilk} &=& {\hat V}_{j1}{\hat V}_{1i}\da~ \delta_{lk}~ k(x_{ik},x_{jk}) 
             - 2 {\hat U}_{i1} {\hat U}\da_{1j}~
\delta_{lk}~ \sqrt{x_{ik}x_{jk}} j(x_{ik},x_{jk}) \nonumber \\
         &&  -\delta_{ij}~ {\hat H}_{k q_L} {\hat H}\da_{q_L l}~ 
k(x_{ik},x_{lk})~.
\eeqa
Here $g_t=m_t/(\sqrt{2}m_W\sin\beta)$ is the top Yukawa 
coupling; $V$ is the CKM matrix; ${\hat V}$ and ${\hat U}$
are the unitary matrices that diagonalize the chargino mass matrix
(~${\hat U}^* M_\chi {\hat V}\da = \mbox{diag}(M_{\chi_1},M_{\chi_2})$~)
and ${\hat H}$ is the one that diagonalizes the up-squark mass matrix
(written in the basis where the $d_L^i-\tu_L^j-\chi_n$ 
coupling is family diagonal and the $d_L^i-\tu_R^j-\chi_n$ 
one is ruled by the CKM matrix). The explicit expressions of 
$k(x,y)$ and $j(x,y)$ can be found in \cite{CI,MWBRS} and, 
as usual, $x_{ij}$ denote ratios of squared masses.

The product of $A^s_{jl}$ and ${\bar A}^b_{ik}$ in (\ref{eq:Zchi})
generates four independent terms, proportional to 
$g^2_t V_{tb}^*V_{ts}$, $g_t V_{ts}$, $g_t V_{tb}^*$ and $1$, 
respectively. As a first approximation we can 
neglect those proportional to $V_{ts}$,
which are clearly suppressed with respect to the SM 
contribution. A further simplification can be obtained 
employing the so-called mass-insertion approximation,
i.e. expanding the up-squark mass matrix around its diagonal.
In this way it can been shown that the potentially 
dominant contribution is the one generated to the first order 
by the $ \ttop_R -\tu^{s}_L $ mixing \cite{LMSS}, namely
\beqa
&&   Z_{sb}^{L}\vert^{\rm RL}_{\rm SUSY}  \, = \,  
- {1\over 8} g_t V_{tb}^* 
~\frac{(M^2_U)_{t_R s_L}}{M^2_{\tu_L}}~
{\hat V}\da_{1j} \Big[ {\hat V}_{j1} {\hat V}\da_{1i} 
k(x_{i u_L}, x_{j u_L}, x_{t_R u_L}) 
\nonumber\\&&\quad -\delta_{ij}  k(x_{i u_L}, x_{t_R u_L}, 1) 
- 2  {\hat U}_{i1} {\hat U}\da_{1j}  \sqrt{x_{i u_L} x_{j u_L}} 
j(x_{i u_L},x_{j u_L}, x_{t_R u_L}) 
\Big] {\hat V}_{i2}~.\quad\quad
\label{eq:WLR}
\eeqa
Notice that, contrary to the $Z^L_{ds}$ case, 
here the CKM factor $V_{tb}^*$ does not imply any 
additional suppression and therefore the  double left-right 
mixing discussed in \cite{CI} represents only a subleading 
correction. In $Z_{sb}^{L}\vert^{\rm RL}_{\rm SUSY} $
the necessary $SU(2)_L$ breaking ($\Delta I_W =1$)
is equally shared by the left-right mixing of the squarks 
and by the chargino-higgsino mixing (shown by the mismatch of 
$\hat V$ indices), carrying both $\Delta I_W =1/2$.

For a numerical evaluation, varying the SUSY parameters 
entering (\ref{eq:WLR}) in the allowed ranges, we find 
\beq
\left| Z_{sb}^{L}\vert^{\rm RL}_{\rm SUSY} \right|   \lsim  0.1 
\left| \frac{(M^2_U)_{t_R s_L}}{M^2_{\tu_L}} \right| = 0.1 
\left| (\delta^U_{RL})_{32}  \right|~,
\label{eq:Zsusynum}
\eeq
in agreement with the results of \cite{LMSS}.
The factor $(\delta^U_{RL})_{32}$, which represents the analog of 
$V_{ts}$ in the SM case, is not very constrained 
at present \cite{LMSS,MWBRS} and can be of $\cO(1)$,
with an arbitrary $CP$-violating phase \cite{LS}.

Eq. (\ref{eq:Zchi}-\ref{eq:Zsusynum}) can simply be extended 
to the $b\to d$ case with the replacement $s\to d$.
Similarly to $(\delta^U_{RL})_{32}$, also $(\delta^U_{RL})_{31}$
is essentially unconstrained at present.

As it can be checked by the detailed analysis of \cite{LMSS}, 
in the interesting limit where the left-right 
mixing of the squarks is the only non-standard 
source of flavour mixing, the $Z$-penguin terms discussed 
above are largely dominant with respect to supersymmetric 
box and $\gamma$-penguin contributions to $b\to s \ell^+\ell^-$.
On the other hand, we note that in processes of the 
type $b\to s q\bar q$ these true penguin terms could 
easily compete in size with the so-called trojan-penguin 
amplitudes discussed in \cite{GKN}.

\subsection{Strong electroweak symmetry breaking}

The natural alternative to low-energy supersymmetry
is the scenario where the Higgs field is not 
elementary and the electroweak symmetry breaking
is generated by some new strong dynamics 
appearing at a scale $\Lambda \sim 1$~TeV. 
Without a detailed knowledge of the new dynamics,
and of the new degrees of freedom associated with it,
a convenient way to describe this scenario 
is obtained by considering the most general effective 
Lagrangian written in terms of fermions and 
gauge fields of the SM, as well as the 
Nambu-Goldstone bosons associated with the spontaneous breaking of 
$SU(2)_L \times U(1)_Y \to U(1)_{\rm em}$  \cite{Longhitano}. 
In this way, imposing the custodial $SU(2)$ symmetry 
on the Nambu-Goldstone boson sector,  
the lowest order terms in the Lagrangian are completely 
determined, corresponding to the SM case in the limit of 
infinite Higgs mass. On the other hand, the effect of the 
new dynamics is encoded in the Wilson coefficients of 
higher-order operators, suppressed 
by appropriate inverse powers of $\Lambda$.

A conservative assumption, usually employed
to reduce the number of free parameters, is that the 
higher-order operators do not involve directly the fermionic sector.
In other words, it is assumed that the new dynamics
involves only the interactions of electroweak 
gauge fields and Nambu-Goldstone bosons \cite{Longhitano}. 
Under this assumption most of the coefficients of the allowed 
dimension-four operators 
(appearing at the next-to-leading order) are strongly 
constrained by electroweak precision data. However, as 
pointed out in \cite{Bernabeu,Burdman_ngb,Burdman_3}, 
some of them naturally escape these bounds and could 
show up in sizable modifications of FCNC amplitudes. 
Interestingly, this happens despite the intrinsic 
flavor-conserving nature of these terms. It occurs 
at the loop level, either via modifications of the 
trilinear gauge-boson couplings \cite{Burdman_3} or via 
corrections to the Nambu-Goldstone boson propagators 
\cite{Bernabeu,Burdman_ngb}. 

Also within this context the FCNC couplings of the $Z$ play 
a special role. As an example, we consider here the 
effect of the anomalous $WWZ$ coupling. Following the 
work of Ref.~\cite{Burdman_3}, this can be written as  
\beqa
  \label{eq:4thZtgc} 
  Z^L_{qb}\vert_{WWZ} &=&  \alpha_3 g^2 V_{tb}^* V_{tq} \frac{3 x_t}{8}
  \log\left(\frac{M_W^2}{\Lambda^2}\right) + \ldots \no\\  
  &\sim & \cO(1) \times V_{tb}^* V_{tq} \frac{g^2 m_t^2}{\Lambda^2} 
\log\left(\frac{M_W^2}{\Lambda^2}\right)~.
\eeqa
where $g$ is the usual $SU(2)_L$ coupling and the dots denote
additional finite terms (i.e. not logarithmically enhanced).
The adimensional coupling $\alpha_3$ 
is one of the unknown coefficients appearing in the 
next-to-leading order Lagrangian
of Ref.~\cite{Longhitano}. This is essentially unconstrained 
by other processes (unless further assumptions are employed) 
and is expect to be of $\cO(M^2_W/\Lambda^2)$ by 
dimensional arguments. The relative shift of 
$Z^L_{qb}$ with respect to the SM case
can thus be up to $50\%$. Interestingly,
the same relative shift would be present in
$Z^L_{ds}$, leading to interesting correlations 
between rare $B$ and $K$ decays \cite{Burdman_3}.
It is worthwhile to point out that this is the only 
non-standard FCNC effect due to anomalous gauge-boson
couplings which is logarithmically 
divergent, which can be taken as an indication of a
particular sensitivity of $Z^L_{ds}$ to the new dynamics
\cite{Burdman_3}.
We finally note that also within this context 
$Z^R_{qb}$ remains unaffected: this is clearly due 
to the chiral nature of the SM gauge group and indeed 
it remains valid also if we consider the effects due to
modified Nambu-Goldstone boson  propagators \cite{Burdman_ngb}.

If the conservative assumption that 
higher-order operators do not involve directly 
the fermionic sector is relaxed, the freedom in 
generating new FCNC effects is clearly enhanced.
The first natural step is to include only
higher-order operators which involve the 
quarks of the third generation, as for instance 
done in \cite{Burd3gen}. However, the most general 
scenario is obtained by considering all generations. 
In this latter option one could generate 
FCNC transitions already at the tree-level \cite{Zhang} and,  
by restricting the attention to the lowest-dimensional  
operators, one would recover the general case
described by Eq.~(\ref{eq:Zsb}).
The predictivity of this scenario is obviously very limited, 
but still, only on dimensional arguments, one 
can conclude that the FCNC couplings of the $Z$ 
could play a very special role. The natural suppression
of FCNC would then suggest 
$Z^{L,R}_{qb} \sim  \cO(m_t^2/\Lambda^2) \times V_{tb}^* V_{tq}$,
leaving open the possibility of $\cO(1)$
corrections with respect to the SM case.

\subsection{Tree-level $Z$-mediated FCNC couplings}
FCNC couplings of the $Z$ can be generated already at the 
tree level in various exotic scenarios. Two popular examples 
discussed in the literature are the models with addition of
non-sequential generations of quarks (see e.g. \cite{GNR} 
and references therein) 
and those with an extra $U(1)$ symmetry (see e.g. \cite{LP} 
and references therein).
In the former case, adding a different number of up- and 
down-type quarks, the pseudo CKM matrix needed to diagonalize 
the charged currents is no more unitary and this leads to 
tree-level FCNC couplings. 
On the other hand, in the case of an extra $U(1)$ symmetry
the FCNC couplings
of the $Z$ are induced by $Z-Z^\prime$ mixing, provided the 
SM quarks have family non-universal charges 
under the new $U(1)$ group. 
Interestingly these two possibilities (i.e. the extra $U(1)$ and 
the non-sequential quarks) are often linked in many 
consistent extensions of the SM \cite{LL}.
Here we will not discuss any of such model in detail.
We simply note, however, that for our purposes these could be well 
described by the effective Lagrangian in (\ref{eq:Zsb}), provided 
the contribution of the $Z^\prime$ exchange is negligible 
or the couplings of the $Z^\prime$ to light charged leptons and 
neutrinos are proportional to the SM ones.

\section{Generalities of exclusive $b\to s\ell^+\ell^-(\nu\bar{\nu})$ decays}
\subsection{Effective Hamiltonian}
\label{sect:heff}
The starting point for the analysis of $b \to s \ell^+\ell^-(\nu\bar{\nu})$ 
transitions is the determination of the low-energy effective
Hamiltonian, obtained by integrating 
out the heavy degrees of freedom of the theory, 
renormalized at a scale $\mu={\cal O}(m_b)$. 
In our framework this can be written as 
\beq
{\cal{H}}_{eff} = - \frac{G_F}{\sqrt{2}}  V_{t s}^\ast  V_{tb}  
\left( \sum_{i=1}^{10} \left[ \c_i  \op_i +
\c_i^{\prime} \op_i^{\prime} \right]
+ \c_L^\nu \op_L^\nu+\c_R^\nu\op_R^\nu \right) \,+\, {\rm h.c.}\; ,
        \label{eq:he}
\eeq
where $\op_i$ denotes the Standard Model basis of operators relevant 
to $b \to s \ell^+ \ell^-$ \cite{BBL} and $\o_i^\prime$ their helicity 
flipped counter parts. In particular, we recall that 
$\op_i\sim(\bar sb)(\bar cc)$, for $i=1 \ldots 6$, 
$Q_8\sim m_b \bar s (\sigma\cdot G) b$, whereas
the only operators with a  tree-level non-vanishing 
matrix element in $b \to s \ell^+ \ell^-$ are given by
\begin{eqnarray}
&\op_7           = \dis\frac{e}{4 \pi^2} \bar{s}_L \sigma_{\mu \nu} 
                   m_b b_R F^{\mu \nu}~,  \qquad\
&\op^\prime_7    = \frac{e}{4 \pi^2} \bar{s}_R \sigma_{\mu \nu} 
                   m_b b_L F^{\mu \nu}~,  
                   \nonumber \\
&\op_9           = \dis\frac{e^2}{4 \pi^2} \bar{s}_L \gamma^\mu b_L 
                   \bar{\ell} \gamma_\mu \ell~, \qquad\qquad
&\op^\prime_9    = \frac{e^2}{4 \pi^2} \bar{s}_R \gamma^\mu b_R 
                   \bar{\ell} \gamma_\mu \ell~, \nonumber \\
&\op_{10}        = \dis\frac{e^2}{4 \pi^2} \bar{s}_L \gamma^\mu b_L 
                   \bar{\ell} \gamma_\mu \gamma_5 \ell~, \qquad\ \ 
&\op^\prime_{10} = \frac{e^2}{4 \pi^2} \bar{s}_R \gamma^\mu b_R 
                   \bar{\ell} \gamma_\mu \gamma_5 \ell~.
\end{eqnarray}
The last two operators in ${\cal{H}}_{eff}$  are defined as
\begin{eqnarray}
\op_{L,R}^\nu = \frac{e^2}{4 \pi^2} \bar{s}_{L,R} \gamma_\mu b_{L,R} 
\bar{\nu} \gamma^\mu (1-\gamma_5) \nu 
\end{eqnarray}
and constitute the complete basis relevant to $b \to s \nu \bar{\nu}$.

Due to the absence of flavour-changing right-handed currents,
within  the Standard Model one has
\beq
C^\prime_{1-10} \vert_{\rm SM} = C_R^\nu \vert_{\rm SM} =0~. 
\eeq
whereas the remaining non-vanishing coefficients are
known at the next-to-leading order \cite{BBL,MU,BB99}. 
The coefficients of $Q_{10}$ and $Q_L^\nu$ are scale independent
and are completely dominated by short-distance dynamics
associated with top quark exchange. Their values are therefore well
approximated by the leading order results, given by 
($\bar m_t(m_t)=166\,{\rm GeV}$)\footnote{
Here and in the following we employ the running ($\overline{MS}$)
mass for the top quark, $\bar m_t(m_t)$. For $b\to s\ell^+\ell^-$ 
the distinction between the pole mass and the running
mass enters, strictly speaking, only beyond the next-to-leading order
we are working in \cite{BMU}. However, the short-distance $\overline{MS}$-mass
is the more appropriate definition for FCNC processes involving virtual
top quarks, and the higher order corrections are generally better
behaved. This is true in particular
for the transitions $b\to s\nu\bar\nu$ and $B_s\to\mu^+\mu^-$, where
the use of the running mass in the known next-to-leading order
expressions is entirely well defined and leads indeed to a small
size of the NLO QCD corrections.}
\beq
C^\nu_L\vert_{\rm SM} = \frac{4B_0(x_t) -
C_0(x_t)}{\sin^2\Theta_W}~= -6.6~, \qquad 
C_{10}\vert_{\rm SM}  = \frac{ B_0(x_t) - 
C_0(x_t)}{\sin^2\Theta_W}~= -4.2~, \label{eq:CiSM}
\eeq
where the contribution proportional to $C_0(x_t)$ is the one induced by 
$Z^L_{sb}\vert_{\rm SM}$ in Eq.~(\ref{eq:SMZsb}) once the $Z$ field has
been integrated out (the full expression for 
$B_0(x)$ can be found in \cite{BBL}). The  difference among the two 
numerical values in (\ref{eq:CiSM}) can be taken as an indication of 
the size of the non-$Z$-induced contributions to these coefficients 
within the SM. On the other hand, 
in the generic  non-standard scenario 
described by $\cL^{Z}_{FC}$ we can write 
\beq
 C^\nu_L -C^\nu_L\vert_{\rm SM} = 
  C_{10} -C_{10}\vert_{\rm SM} =  
 - \frac{  Z^L_{bs}- Z^L_{bs}\vert_{\rm SM} }{V_{ts}^*V_{tb} \sin^2\Theta_W}~,
 \qquad 
C^\nu_R = C^\prime_{10} = - \frac{  Z^R_{bs}  }{V_{ts}^*V_{tb}
 \sin^2\Theta_W} ~. \label{eq:nonstandardCi}
\eeq

In principle the coefficients $C_9$ and $C_9^\prime$ are 
also sensitive to $Z^L_{bs}$ and $Z^R_{bs}$. In this case, 
however, the contribution of $\cL^{Z}_{FC}$ 
is suppressed by the smallness of the vector coupling of the 
$Z$ to charged leptons ($|a_V^e/a_A^e| = |4 \sin^2\Theta_W -1| 
\simeq 0.08$)
and as a first approximation can be neglected.
Given the above considerations, we will assume
in the following that all the Wilson coefficients but those 
in (\ref{eq:nonstandardCi}) coincide with their SM expressions. 

\subsection{Kinematics and form factors}
\label{sect:OKincl}

In the following sections we shall discuss integrated observables 
and distributions in the invariant mass of the dilepton system, $q^2$, 
for the three-body decays 
$B\to H  \ell \bar \ell$, with $H=K$, $K^*$ and 
$\ell=\mu,\ \nu$. The kinematical range of $q^2$ is 
given  by $4 m_\ell^2 \simeq 0\leq q^2 \leq (m_B-m_H)^2$.
In the neutrino case  $q^2$ is not directly measurable but is 
related to the kaon energy in the $B$ meson rest frame, 
varying in the interval 
$m_H\leq E_H\leq (m^2_B+m^2_H)/(2 m_B)$
by the relation 
$q^2 = m^2_B+m^2_H-2 m_B E_H$.
For convenience we define also the dimensionless variables
$s =q^2/m_B^2$ and $r_H =m_H^2/m^2_B$, and the function
\beq
\lambda_H(s) =  1 + r_H^2 + s^2 - 2s - 2r_H - 2r_Hs~.
\eeq
%
%
In the case  $H=K$
the hadronic matrix elements needed for our analysis can be written as 
\beqa
\langle\bar K(\pk)|\bar s\gamma_\mu b|\bar B(p)\rangle &=&
f_+(q^2)(p+\pk)_\mu+f_-(q^2)q_\mu~, \label{fpmdef} \\
q^\nu \langle\bar K(\pk)|\bar s\sigma_{\mu\nu}b|\bar B(p)\rangle &=&
i\frac{f_T(q^2)}{m_B+m_K}\left[q^2 (p+\pk)_\mu-(m_B^2-m_K^2)q_\mu\right]~,
\label{atdef}
\eeqa
where $q^\mu=p^\mu-\pk^\mu$.
Up to small isospin breaking effects, which we shall neglect,
the same set of form factors describes both charged ($B^- \to K^-$)
and neutral ($\bar B^0 \to \bar K^0$) transitions.
Similarly, in the case $H=K^*$ we can write ($\epsilon^{0123}=+1$) 
\beqa
&& \langle\bar \Ks(\pk,\eps) |\bar s\gamma_\mu\gamma_5 b|\bar B(p)\rangle ~=~
2 m_\Ks A_0(q^2)\frac{\eps^*\cdot q}{q^2}q_\mu + (m_B + m_\Ks )
A_1(q^2)\left[\eps^*_\mu-\frac{\eps^*\cdot q}{q^2}q_\mu\right]\no\\
&&\qquad -A_2(q^2)\frac{\eps^*\cdot q}{m_B + m_\Ks}
\left[(p+\pk)_\mu-\frac{m^2_B - m^2_\Ks}{q^2}q_\mu\right]~, \\
&& \langle\bar \Ks(\pk,\eps) |\bar s\gamma_\mu b|\bar B(p)\rangle ~=~
 i\frac{2 V(q^2)}{m_B + m_\Ks}  
\epsilon_{\mu\nu\rho\sigma}\eps^{*\nu} p^\rho \pk^\sigma~, \\
&&  q^\nu 
\langle\bar \Ks(\pk,\eps)|\bar s\sigma_{\mu\nu}(1+\gamma_5)b|\bar B(p)\rangle 
 ~=~ -2T_1(q^2) \epsilon_{\mu\nu\rho\sigma}\eps^{*\nu} p^\rho \pk^\sigma \no \\
&&\qquad -i T_2(q^2)
\left[ \eps^*_\mu(m_B^2 -m_\Ks^2) -(\eps^*\cdot q)(p+\pk)_\mu\right] 
   -i T_3(q^2)(\eps^*\cdot q) 
\left[ q_\mu - \frac{q^2 (p+\pk)_\mu }{m_B^2 -m_\Ks^2}\right]\qquad
\eeqa
Here we have used the phase conventions of \cite{LEET}. In particular,
all form factors are real and positive. We remark that the large-energy
limit discussed in \cite{LEET} is especially useful to fix
the relative sign of the various form factors in a model
independent way.

\noindent
The form factors $f_T$, $T_1$, $T_2$ and $T_3$ depend on the
renormalization scale, which here and in the following is understood
to be $\mu=m_b$. There is no need to further specify the
renormalization scheme for the tensor operator
$\bar s\sigma_{\mu\nu}(1+\gamma_5)b$, since the issue of a non-trivial
scheme dependence enters only beyond the next-to-leading
logarithmic approximation in $b\to s\ell^+\ell^-$

For the numerical evaluations of $f_i(q^2)$, $A_i(q^2)$, $T_i(q^2)$
and $V(q^2)$
 we refer to the recent 
analysis of Ref.~\cite{ABHH99}, performed in the framework 
of light-cone sum rules.

\section{$B\to (K,K^*) \nu \bar{\nu}$}
\label{sect:nnmodes}
{}From a theoretical point of view the neutrino channels are certainly 
much cleaner compared to the charged lepton ones due to the absence of 
long-distance effects of electromagnetic origin. Moreover
the smaller number of operators involved (only two) simplifies  
their description. Finally the branching fractions are
enhanced by the summation over the three neutrino flavours. 
All these virtues, however, are partially compensated 
by the difficult experimental signature.

\subsection{$B\to K \nu \bar{\nu}$}
\begin{figure}[t]
\vskip 0.0truein
\centerline{\epsfysize=3.5in   
{\epsffile{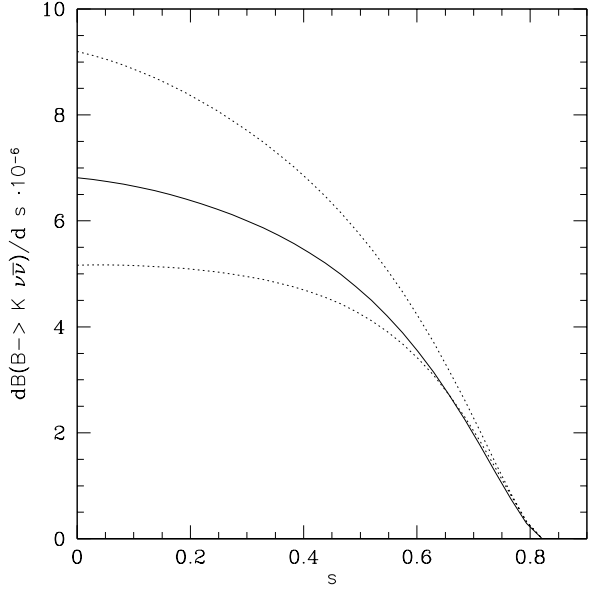}}}
\vskip 0.0truein
\caption[]{ \it Dilepton invariant mass distribution for 
$\Br(B\to K \nu \bar{\nu})$ within the SM. The three lines
correspond to the central, minimal and maximal values  
of $f_+(s)$ from \cite{ABHH99}. }
\label{fig:bknunu}
\end{figure}

The dilepton spectrum of this mode is 
particularly simple and is sensitive only to 
the combination $|\c_L^\nu +\c_R^\nu|$ \cite{Colangelo}:
\beq
  \frac{d \Gamma (B \to K \nu \bar{\nu})}{d s}  =  
  \frac{G_F^2  \alpha^2  m_B^5}{256 \pi^5} 
      \left| V_{ts}^\ast  V_{tb} \right|^2  \lambda_K^{3/2}(s) 
f^2_+(s) |\c_L^\nu +\c_R^\nu|^2 
\label{eq:dBKnunu}
\eeq
The differential branching ratio computed within the SM is plotted in 
Fig.~\ref{fig:bknunu}, showing the uncertainty due to the 
form factors. Note that in the neutral modes the strangeness eigenstates
of the kaons do not coincide with the mass eigenstates, which
are experimentally detected. Therefore, neglecting isospin-breaking 
and $\Delta S=2$ $CP$-violating effects, we can write
\beq
  \Gamma(B \to K \nu \bar{\nu}) \equiv \Gamma(B^+ \to K^+ \nu \bar{\nu})=
2 \Gamma(B^0 \to K_{L,S} \nu \bar{\nu})~. 
\eeq
The absence of absorptive final-state interactions in this 
process also leads to 
$\Gamma(B \to K \nu \bar{\nu})=\Gamma(\bar B \to \bar K \nu \bar{\nu})$,
preventing the observation of any direct-$CP$-violating effect.

Integrating Eq.~(\ref{eq:dBKnunu}) over the full range of $s$ leads to
\beqa
\label{eq:BRKnunu}
{\cal{B}}(B\to K \nu \bar{\nu})&=&(3.8^{+1.2}_{-0.6}) \times 10^{-6}
~\left|\frac{\c_L^\nu +\c_R^\nu }{\c_L\vert_{SM}^\nu} \right|^2\no\\
&\approx&  4\times 10^{-6}~\left| 1- \frac{  (Z^L_{bs}-Z^L_{bs}|_{SM}) 
+ Z^R_{bs} }{ 0.06 }
\right|^2~, \label{eq:Bbkpnn}
\eeqa
where the error in the first equality is due to the uncertainty 
in the form factors and the second relation 
has been obtained by means of Eq.~(\ref{eq:nonstandardCi}).
Given the constraint (\ref{eq:Zsblim}), without further 
assumptions we find
$\Br(B\to K \nu \bar{\nu})\lsim 5\times 10^{-5}$.
This bound sets the level below which an experimental 
constraint on this mode starts to provide significant information. 
On the other hand, in most of the scenarios discussed in 
Section~\ref{sect:3}, where  $Z^R_{bs}= 0$ and  $|Z^L_{bs}|\lsim 0.1$, 
we find 
\beq
\Br(B\to K \nu \bar{\nu})\lsim 2 \times 10^{-5}~.
\label{eq:nnb1}
\eeq

If the experimental sensitivity on $\Br(B\to K \nu \bar{\nu})$ 
reached the $10^{-6}$ level, 
then the uncertainty due the form factors would prevent 
a precise extraction of $|\c_L^\nu +\c_R^\nu|$ from 
(\ref{eq:Bbkpnn}). This problem can be substantially reduced 
by relating the differential distribution 
of $B\to K \nu \bar{\nu}$ to the one of $B \to \pi e \nu_e$
\cite{FG,AlievKim}:
\beq
\frac{d \Gamma(B \to K \nu \bar{\nu})/ds }
{d \Gamma(B^0 \to \pi^- e^+ \nu_e) /ds  }=
\frac{3 \alpha^2}{4 \pi^2} \left|\frac{ V_{ts}^\ast  V_{tb}}{V_{ub}}\right|^2
\left(\frac{\lambda_K(s)}{\lambda_\pi(s)}\right)^{3/2}
\left\vert \frac{ f_+^K(s)}{f_+^\pi(s)} \right\vert^2 |\c_L^\nu +\c_R^\nu|^2~.
\label{eq:Bknn_r}
\eeq
Indeed $f_+^K(s)$ and $f_+^\pi(s)$ coincide up to 
$SU(3)$ breaking effects, which are expected to be 
small, especially far from the endpoint region.
An additional uncertainty in (\ref{eq:Bknn_r}) is induced by 
the CKM ratio $| V_{ts}^\ast  V_{tb}|^2/|V_{ub}|^2 $ which,
however, can independently  be determined from other processes.

\subsection{$B\to \Ks \nu \bar{\nu}$}
\begin{figure}[t]
\vskip 0.0truein
\centerline{\epsfysize=3.5in   
{\epsffile{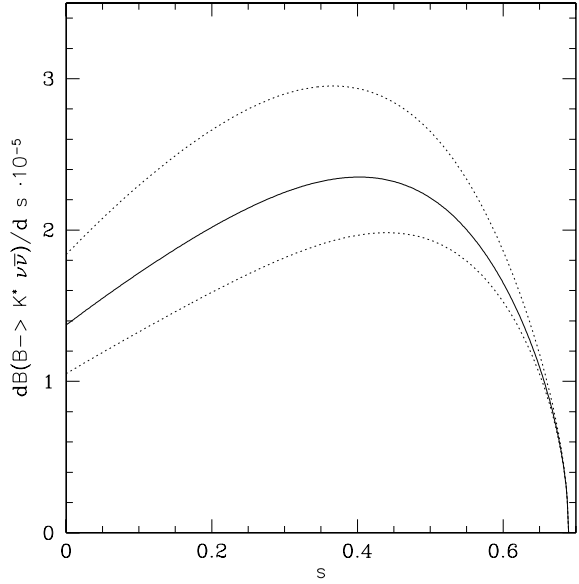}}}
\vskip 0.0truein
\caption[]{\it Dilepton invariant mass distribution for 
$\Br(B\to \Ks \nu \bar{\nu})$ within the SM. The three lines
correspond to the central, minimal and maximal values,  
as obtained by varying the form factors 
within the ranges quoted in \cite{ABHH99}.}
\label{fig:bkstnunu}
\end{figure}

The dilepton invariant mass spectrum of $B\to \Ks \nu \bar{\nu}$ decays is 
sensitive to both combinations $|C_L^\nu - C_R^\nu|$ 
and  $|C_L^\nu + C_R^\nu|$ \cite{Colangelo,KKM}:
\beqa
  \frac{d \Gamma (B\to \Ks \nu \bar{\nu})}{d s} & = & 
    \frac{G_F^2  \alpha^2  m_B^5}{1024 \pi^5} 
    \left| V_{t s}^\ast  V_{tb} \right|^2  \lambda_\Ks^{1/2}(s)
    \Bigg\{ \frac{ 8 s \lambda_\Ks(s)V^2(s) }{(1+\sqrt{r_\Ks})^2} 
    \left|C_L^\nu +C_R^\nu\right|^2 \Bigg.  \no\\
&& +\frac{1}{r_\Ks} \Bigg[ (1+\sqrt{r_\Ks})^2 \left( \lambda_\Ks(s)+
   12 r_\Ks s \right) A_1^2(s) 
  +\frac{\lambda_\Ks^2(s)A_2^2(s)}{(1+\sqrt{r_\Ks})^2} \Bigg. \no\\
&& \Bigg.\Bigg.  \qquad\qquad - 2 \lambda_\Ks(s) (1-r_\Ks-s) 
  A_1(s) A_2(s) \Bigg] \left| C_L^\nu -C_R^\nu \right|^2  \Bigg\}~.\ 
   \label{eq:dBKstnunu}
\end{eqnarray}
The  branching fraction obtained within the SM 
is shown in Fig.~\ref{fig:bkstnunu}.

Integrating  Eq.~(\ref{eq:dBKstnunu}) over the full range of $s$ leads to
\begin{eqnarray}
\Br(B\to \Ks \nu \bar{\nu})
&=& (2.4^{+1.0}_{-0.5}) \times 10^{-6} 
~\left|\frac{\c_L^\nu +\c_R^\nu }{\c_L\vert_{SM}^\nu} \right|^2
+ (1.1^{+0.3}_{-0.2}) \times 10^{-5}  
~\left|\frac{\c_L^\nu - \c_R^\nu }{\c_L\vert_{SM}^\nu} \right|^2~, 
\label{eq:BRKstnunu} \\
\Br(B\to \Ks \nu \bar{\nu})\Big|_{\rm SM} &=& (1.3^{+0.4}_{-0.3}) \times 10^{-5}~.
\label{eq:BRKstnunuSM}
\end{eqnarray}
Similarly to the case of $\Br(B\to K \nu \bar{\nu})$, 
the bound (\ref{eq:Zsblim}) leaves open the possibility of
enhancements of  $\Br(B\to \Ks \nu \bar{\nu})$
up to one order of magnitude 
with respect to the SM case. 
Whereas if  $Z^R_{bs}=0$ and  $|Z^L_{bs}|\lsim 0.1$,
we find the constraint 
\beq
\Br(B\to \Ks \nu \bar{\nu}) \lsim  10^{-4}~,
\label{eq:nnb2}
\eeq
which is almost one order of magnitude below
the present experimental 
sensitivity \cite{delphi96}.

A reduction of the error induced by the poor knowledge of the 
form factors can be 
obtained by normalizing the dilepton distributions of
$B\to K^* \nu \bar{\nu}$ to the one of
$B \to \rho e \nu_e$ \cite{LWise,AlievKim}. 
This is particularly effective in the limit $s\to 0$, 
where the contribution proportional to 
$|C_L^\nu + C_R^\nu|$ (vector current) drops out:
\begin{eqnarray}
\left. \frac{d \Gamma(B \to K^* \nu \bar{\nu})/ds }{d 
\Gamma(B^0 \to \rho^- e^+ \nu_e )/ds }\right|_{s=0} &=&
\frac{3 \alpha^2}{4 \pi^2} \left|\frac{ V_{ts}^\ast  V_{tb}}{V_{ub}}\right|^2
\left(\frac{1-r_\Ks}{1-r_\rho}\right)^3 
\frac{r_\rho}{r_\Ks} \left|C_L^\nu -C_R^\nu\right|^2 \nonumber \\
&& \times
\left\vert \frac{A_1^{K^*}(0)(1+\sqrt{r_\Ks})-
A_2^{K^*}(0)(1-r_\Ks)/ (1+\sqrt{r_\Ks}) }
{A_1^{\rho}(0)(1+\sqrt{r_\rho})-
A_2^{\rho}(0)(1-r_\rho)/(1+\sqrt{r_\rho}) } 
\right\vert^2\;
\label{eq:BKsnn_r}
\end{eqnarray}
Similarly to the ratio $f_+^K(s)/f_+^\pi(s)$ in (\ref{eq:Bknn_r}),
also the last term in (\ref{eq:BKsnn_r}) is equal to one 
up to $SU(3)$-breaking corrections.

\section{$B\to (K,K^*) \ell^+\ell^-$}

The possibility to detect the leptons not 
only provides a clear experimental signature
for $B\to (K,K^*) \ell^+\ell^-$ decays,
it also allows to consider interesting observables
in addition to the decay distribution,
like the forward-backward asymmetry.
Moreover, the non-vanishing absorptive 
contributions lead to potentially large
direct-$CP$-violating effects. 

The problem of these modes is the uncertainty 
in the non-perturbative contributions generated by the 
operators $Q_{1-6}$ in ${\cal{H}}_{eff}$.
Indeed these induce transitions of the type
$b\to s (c\bar c) \to s \ell^+ \ell^-$ that
can be handled in perturbation theory only 
within specific regions of the  dilepton spectrum.

In the following we shall restrict our attention to the 
transitions with a $\mu^+ \mu^-$ pair in the final state,
which have the clearest experimental signature,
however the whole discussion is equally applicable 
to the $e^+ e^-$ case.

\subsection{Non-perturbative $c \bar c$ corrections and $\cne$}
\label{sect:cne}
In the kinematic  region of large dilepton invariant 
mass, above the $\Psi^\prime$ peak,
the light quark fields  ($u,~d,~s,~c$) appearing in
${\cal{H}}_{eff}$ may be integrated out explicitly since 
they enter loop diagrams with a hard external scale
($q^2 \sim m^2_b$) \cite{LWise,BI}.
The endpoint effective Hamiltonian thus derived, 
valid at the  next-to-leading order in QCD, 
can be obtained from the one in (\ref{eq:he}) setting to 
zero the coefficients of $Q_{1-6}$ 
and replacing  $C_9$ with 
\beq
C_9^{\rm EP}(s) = C_9 + h\left(\frac{m_c}{m_b},\frac{m^2_B}{m^2_b}s \right) 
(3C_1 +C_2)  + \cO\left( C_{3-6} \right )~, \label{eq:C9EP}
\eeq
where the function $h(x,y)$ and the numerically 
small $\cO\left( C_{3-6} \right )$ terms can be found in \cite{BMM}
(we recall that to the next-to-leading order accuracy, only the
leading order values of $C_{1-6}$ are need in $C_9^{\rm EP}$).
Note that the coefficient function in (\ref{eq:C9EP}) differs from the
effective coupling of $Q_9$ usually introduced 
to describe inclusive decays \cite{BBL}, since it does not include
the QCD correction to the matrix element of the
${\bar s}_L \gamma^\mu b_L$ current. Indeed the latter has to be 
included in the corresponding hadronic matrix elements, 
assuming they are computed in 
full QCD and appropriately normalized at 
$\mu={\cal O}(m_b)$. 

In the region of large $q^2$ one still expects non-perturbative 
corrections induced by intermediate $c \bar c$ states.
Although in principle power suppressed ($\sim \Lambda_{QCD}/m_b$), 
locally these are likely to produce sizable modifications
to the dilepton spectrum. 
The relative importance of these non-perturbative effects, however, can be 
diminished by integrating over sufficiently large ranges of $q^2$.

\begin{figure}[t]
\vskip 0.0truein
\centerline{\epsfysize=3.5in   
{\epsffile{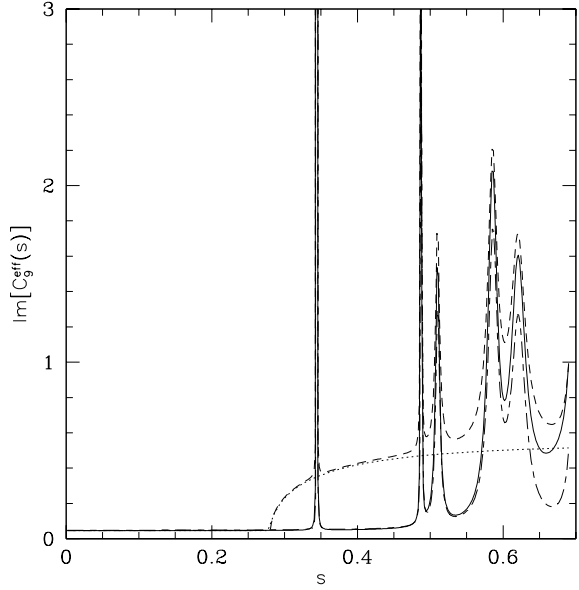}}}
\vskip 0.0truein
\caption[]{ \it The imaginary part of $\cne$ as a function of $s$:
$\Im \cne(s) = \Im C_9^{\rm EP}(s)$ as in (\ref{eq:C9EP}) (dotted);
KS prescription \cite{ks96} (solid); Ref.~\cite{lsw98} (dot-dashed). 
For comparison we have also included the approach of
Ref.~\cite{amm91} (dashed), where
Breit-Wigner resonances are naively
added to the partonic calculation. (This procedure is disfavoured
since it has a manifest problem of double-counting.)}
\label{fig:imC9}
\end{figure}

Far from the endpoint region it is not possible, in principle,
to safely integrate out the light quark fields in ${\cal{H}}_{eff}$ 
and one should estimate separately the matrix elements of $Q_{1-6}$. 
In general this is a very complicated task that 
has so far been treated only with the help of some non-rigorous 
simplifying assumptions. For instance, assuming 
that the matrix elements of $Q_{1-6}$ can be factorized as
\beq
\langle H \mu^+\mu^- | Q_i | \bar B \rangle 
\propto  \langle H | {\bar s}_L \gamma^\mu b_L  | \bar B \rangle 
\times  \langle \mu^+\mu^- | {\bar c} \gamma^\mu c  | 0 \rangle~,
\label{eq:fact_ass}
\eeq
one can employ the Kr{\"u}ger-Sehgal (KS) approach \cite{ks96}
and estimate $\langle \mu^+\mu^- | {\bar c} \gamma^\mu c  | 0 \rangle$
by means of $\sigma(e^+ e^- \to c\bar c)$ data. 
This approach has the advantage of avoiding 
double-counting and to provide 
a rigorous non-perturbative estimate of  
$\langle \mu^+\mu^- | {\bar c} \gamma^\mu c  | 0 \rangle$.
Other recipes to evaluate the contributions of 
$\langle Q_{1-6}\rangle$ can be found e.g. in
\cite{amm91} and  \cite{lsw98}. 
In all cases, in analogy with (\ref{eq:C9EP}), 
these contributions are encoded via an effective 
coupling for the operator $Q_9$ of the type 
\begin{equation}
\cne(s) = C_9 + Y(s)~.
\label{eqn:c9eff}
\end{equation}
Due to the real intermediate $c \bar c$ states,
$Y(s)$ develops an imaginary part
that plays a crucial role in determining the 
size of direct-$CP$-violating observables.
A comparison of the different approaches to 
compute $\Im\cne(s)$ is shown in
Fig.~\ref{fig:imC9}.

In the following we shall compare results 
obtained by identifying $C_9^{\rm eff}(s)$ 
with $C_9^{\rm EP}(s)$
or, alternatively, by employing the KS approach.

\subsection{Branching ratios and dilepton spectra}
\label{sect:BRKpll}
Neglecting the lepton mass, the $q^2$  
distributions of $\bar B\to \bar K \mu^+ \mu^-$ 
and $\bar B\to \bar \Ks \mu^+ \mu^-$ decays, computed 
with the effective Hamiltonian of Section~\ref{sect:heff},
can be written as
\beqa
\frac{d \Gamma (\bar B \to \bar K  \mu^+ \mu^- )}{d s} &=& 
\frac{G^2_F \alpha^2 m^5_B}{1536 \pi^5}
|V_{tb}V_{ts}|^2 \lambda_K^{3/2}(s) ~ \Bigg\{ 
f^2_+(s)\left(|C_9^{\rm eff}(s) |^2 + |C_{10}+C^\prime_{10}|^2 \right) 
\Bigg. \no\\ 
&&\quad +\frac{4 m^2_b f^2_T(s)}{(m_B + m_K)^2}|C_7|^2
+ \frac{4 m_b f_T(s)f_+(s)}{m_B +m_K} \Re\left(
 C_9^{\rm eff}(s)C^*_7\right) \Bigg\}, 
\eeqa
\beqa
&&\frac{d \Gamma (\bar B \to \bar \Ks  \mu^+ \mu^- )}{d s} =
\left. \frac{d \Gamma (\bar B \to \bar \Ks  \mu^+ \mu^- 
)}{d s}\right|_{\rm SM} \no\\
&&\qquad +\frac{G^2_F \alpha^2 m^5_B}{1024 \pi^5}
|V_{tb}V_{ts}|^2 \lambda_\Ks^{1/2}(s) 
\Bigg\{ \frac{4 s\lambda_\Ks(s) V^2(s) }{3 (1+\sqrt{r_\Ks})^2}
\left( |C_{10}+C^\prime_{10}|^2 - |C_{10}|_{\rm SM}|^2 \right)
\Bigg. \no\\
&&\qquad\qquad +\left[\frac{\lambda_\Ks(s)+ 12 r_\Ks s }{6r_\Ks}  
(1+\sqrt{r_\Ks})^2 A^2_1(s)
 -\frac{\lambda_\Ks(s)}{3r_\Ks} (1-r_\Ks-s) A_1(s) A_2(s) \right. \no\\ 
&&\qquad\qquad\quad\ \Bigg. \left. 
+ \frac{\lambda^2_\Ks(s) A_2^2(s) }{6r_\Ks (1+\sqrt{r_\Ks})^2} \right] 
  \left( |C_{10} -C^\prime_{10}|^2 - |C_{10}|_{\rm SM}|^2 \right)
\Bigg\}~.\label{eq:dBrKmm}
\eeqa
The SM expression of  
$d \Gamma (\bar B \to \bar \Ks  \mu^+ \mu^- )/ds$ 
is given by 


\beqa
\frac{d\Gamma(\bar B\to\bar\Ks\mu^+\mu^-)}{d s} &=& 
    \frac{G_F^2  \alpha^2  m_B^5}{1024 \pi^5} 
    \left| V_{t s}^\ast  V_{tb} \right|^2  \lambda_\Ks^{1/2}(s) \\
&&\times\Bigg\{ R_9\left(|C^{\rm eff}_9(s)|^2+|C_{10}|^2\right)+
    R_7\frac{m^2_b}{m^2_B}|C_7|^2+
   R_{97}\frac{m_b}{m_B} {\rm Re}C_9^{\rm eff}(s)C^*_7 \Bigg\}~,\ 
   \label{eq:dBKstmmSM} \no
\end{eqnarray}
where
\begin{eqnarray}\label{r9def}
R_9 &=& \frac{4s\lambda_\Ks (s) V^2(s)}{3(1+\sqrt{r_\Ks})^2}+
\frac{(1+\sqrt{r_\Ks})^2}{6r_\Ks}(\lambda_\Ks (s)+12 r_\Ks s)A^2_1(s)
+\frac{\lambda^2_\Ks (s)}{6r_\Ks}\frac{A^2_2(s)}{(1+\sqrt{r_\Ks})^2}\no\\
&&\quad -\frac{\lambda_\Ks (s)(1-r_\Ks-s)}{3r_\Ks}A_1(s) A_2(s) 
\end{eqnarray}

\begin{eqnarray}\label{r7def}
R_7 &=& \frac{16\lambda_\Ks (s) T^2_1(s)}{3s}+
\frac{2(1-r_\Ks)^2}{3r_\Ks s^2}(\lambda_\Ks (s)+12 r_\Ks s)T^2_2(s)
+\frac{2\lambda^2_\Ks (s)}{3r_\Ks(1-r_\Ks)^2} T^2_4(s)\no\\
&&\quad -\frac{4\lambda_\Ks (s)(1-r_\Ks-s)}{3r_\Ks s}T_2(s) T_4(s) 
\end{eqnarray}

\begin{eqnarray}\label{r97def}
R_{97} &=& \frac{16\lambda_\Ks (s) V(s) T_1(s)}{3(1+\sqrt{r_\Ks})}+
\frac{2(1-r_\Ks)(1+\sqrt{r_\Ks})}{3r_\Ks s}
(\lambda_\Ks (s)+12 r_\Ks s)A_1(s) T_2(s) \no\\
&&\quad +\frac{2\lambda^2_\Ks (s)(1-\sqrt{r_\Ks})}{3r_\Ks(1-r_\Ks)^2}
A_2(s)T_4(s)\no\\
&&\quad -\frac{2\lambda_\Ks (s)(1-r_\Ks-s)}{3r_\Ks}
\left(\frac{1-\sqrt{r_\Ks}}{s}A_2(s) T_2(s)+
\frac{1}{1-\sqrt{r_\Ks}}A_1(s) T_4(s)\right) 
\end{eqnarray}

and we have defined
\begin{equation}\label{t4def}
T_4(s)\equiv T_3(s)+\frac{1-r_\Ks}{s} T_2(s)
\end{equation}

Here we have again neglected the lepton mass,
which is an excellent approximation for $\ell=e$, $\mu$
if $s\gg 4m^2_\ell/m^2_B$.
The full $m_\ell$ dependence can be found for instance
in \cite{ABHH99}.


As it can be noticed, the coefficients $C_{10}$
and $C^\prime_{10}$, which could have a potentially large
$CP$-violating phase induced by $Z^{L,R}_{bs}$,
do not interfere with $C_9^{\rm eff}(s)$, which has a 
non-vanishing $CP$-conserving phase. As a consequence, 
similarly to the SM case, also within our generic 
non-standard scenario we do not expect to observe 
any sizable (i.e. above the $10^{-2}$ level)  
$CP$ asymmetry in the dilepton invariant mass 
distribution of both decay modes. In the remaining 
part of this subsection we will therefore not 
distinguish between $B$ and 
$\bar B$ states.

The integration over the full range of $s$ 
with $C_9^{\rm eff}(s)\equiv C_9^{\rm EP}(s)$ 
(non-resonant branching ratio) and 
the SM Wilson coefficients leads to 
$\Br( B \to \Ks  \mu^+ \mu^-)^{\rm n.r.}|_{\rm SM} =
1.9^{+0.5}_{-0.3}\times 10^{-6}$
and $\Br( B \to K \mu^+ \mu^-)^{\rm n.r.}|_{\rm SM} =
5.7^{+1.6}_{-1.0}\times 10^{-7}$ \cite{ABHH99},
where the error is mainly determined by the 
uncertainty on the form factors. 
Interestingly
$\Br( B \to  \Ks  \mu^+ \mu^-)^{\rm n.r.}|_{\rm SM}$
is quite close to the experimental limit 
\beq
\label{eq:CDFbounds}
\Br(B^0 \to K^{*0} \mu^{+}  \mu^{-})^{\rm n.r.} < 4.0 \times 10^{-6}
\eeq 
recently obtained by CDF \cite{CDF}, whereas
for $\Br(B\to K \mu^+\mu^-)^{\rm n.r.}$ the best bound-to-SM ratio 
is around 9 \cite{CDF}. Thus the $K^*$ mode 
provides a powerful tool to constraint $|C_{10}|$ and 
$|C_{10}^\prime|$, or $|Z^{L,R}_{bs}|$, via the relation
\beqa
\Br( B \to  \Ks  \mu^+ \mu^-)^{\rm n.r.} &=&
 \Br( B \to \Ks  \mu^+ \mu^-)^{\rm n.r.}|_{\rm SM} \no \\
&& + \left(4.1^{+1.0}_{-0.7}\right)\times 10^{-8} 
 \left( |C_{10} -C^\prime_{10}|^2 - |C_{10}|_{\rm SM}|^2 \right) \no\\
&& +\left(0.9^{+0.4}_{-0.2}\right)\times 10^{-8} 
\left( |C_{10} + C^\prime_{10}|^2 - |C_{10}|_{\rm SM}|^2 \right)~,
\label{eq:bound1}
\eeqa
obtained by integrating (\ref{eq:dBrKmm}).
Using the bound (\ref{eq:CDFbounds}) and setting $C_{10}^\prime=0$
we recover the result of \cite{ABHH99}  $|C_{10}| \lsim 10$, 
that in turn implies
\beq
 \label{eq:ZsblimCDF}
   \left| Z^L_{bs} \right| \lsim 0.10~ .
\eeq
Note that, since $C_{10}$ is basically dominated by the $Z$
penguin already within the SM, the maximal allowed value for 
$| Z^L_{bs} |$ is to a good approximation 
independent of the sign of $Z^L_{bs}$.
On the other hand, if we allow 
also $C_{10}^\prime$ to be different 
from zero we find the relation
\beq
\label{eq:ctctpbound}
|C_{10}|^2+|C_{10}^\prime|^2 
- (1.25 \pm 0.05)\times \Re\left(C_{10}^* C_{10}^\prime\right)
\lsim 100~,
\eeq
where the coefficient 
of $\Re(C_{10}^* C_{10}^\prime)$
is quite stable with respect to variations of the 
form factors.
Varying arg$(C_{10}/C_{10}^\prime)$ over $2\pi$
we find $|\ct|$, $|\ct^\prime| \lsim 13$, leading to
\beq
 \label{eq:ZsblimCDFLR}
   \left| Z^{L,R}_{bs} \right| \lsim 0.13~.
\eeq

Due to the uncertainties in the form factors and 
the assumptions on the non-perturbative non-resonant 
contributions, the bounds derived from Eq.~(\ref{eq:bound1}) 
could appear less clean, from a theoretical point of view,
than those derived from the inclusive rates. 
We stress, however, that even doubling the errors on
the form factors the constraints in (\ref{eq:ZsblimCDF})
and (\ref{eq:ZsblimCDFLR}) do not 
increase by more than $10\%$. 

Though still at the border of most of the model 
predictions discussed in Section~\ref{sect:3},
the bound (\ref{eq:ZsblimCDF}) starts to 
provide a significant information.
For instance, it strengthens the model-independent 
character of the bounds (\ref{eq:nnb1})
and (\ref{eq:nnb2}) for the neutrino modes.
As already discussed in Section~\ref{sect:nnmodes},
if the experiments reached the SM sensitivity on 
$B \to  \Ks  \mu^+ \mu^-$, 
more precise information on $C_{10}$ and $C^\prime_{10}$ 
could be obtained by relating the from factors of 
this mode to those of its $SU(3)$ partner 
$B \to  \rho e \nu_e$.

\subsection{Forward-backward asymmetry in 
$B\to K^*\mu^+\mu^-$}
As anticipated, the possibility of detecting the leptons in 
the final state allows us to study interesting asymmetries 
in the decay distribution of $B\to H  \mu^+\mu^-$ modes. 
The (lepton) forward-backward asymmetry of 
$\bar B\to \bar K^* \mu^+\mu^-$ can be defined as 
\beq
\label{eq:asdef}
\cA^{(\bar B)}_{FB}(s)=\frac{1}{d\Gamma(\bar B\to \bar K^* \mu^+\mu^- )/ds}
  \int_{-1}^1 d\cos\theta ~
 \frac{d^2 \Gamma(\bar B\to \bar K^* \mu^+\mu^- )}{d s~ d\cos\theta}
\mbox{sgn}(\cos\theta)~,
\eeq
where $\theta$ is the angle between the momenta of 
$\mu^+$ and $\bar B$ in the dilepton center-of-mass frame. 
Given the vector or axial-vector structure of the 
leptonic current generated by ${\cal{H}}_{eff}$,  
this asymmetry can be different from zero only if 
the final hadronic system has a non-vanishing angular 
momentum and therefore it is identically zero in the case of
$B(\bar B)\to K(\bar K) \mu^+\mu^-$.

The explicit expression for $\cA^{(\bar B)}_{FB}(s)$
in terms of Wilson coefficients and form factors 
can be written as 
\beqa
  \cA^{(\bar B)}_{FB}(s) &=& 
  - \frac{G_F^2  \alpha^2  m_B^5 \left| V_{ts}^\ast  V_{tb} \right|^2 
      }{256 \pi^5 d\Gamma(\bar B\to \bar K^* \mu^+\mu^- )/ds} 
     ~  \lambda_\Ks(s) \left| V(s) A_1(s) \right|    \no \\
& & \times  {\rm Re}\left\{  \ct^* \left[ s~\cne(s) 
    + \alpha_+(s) \frac{m_b C_7}{m_B}  
    + \alpha_-(s) \frac{m_b  C_7 \ct^{\prime*} }{m_B \ct^* } 
  \right] \right\}~,  
  \label{eq:dfbabvllex}
\eeqa
where
\beq
\label{eq:alpahpm}
 \alpha_\pm(s) = \frac{T_2(s)}{A_1(s)}(1-\sqrt{r_\Ks}) \pm 
                 \frac{T_1(s)}{V(s)}(1+\sqrt{r_\Ks})~
\eeq
and we have used the model-independent relation between the signs 
of $V(s)$ and $A_1(s)$, discussed in Section~\ref{sect:OKincl},
to elucidate the overall sign of $\cA^{(\bar B)}_{FB}(s)$.

The ratios of form factors in (\ref{eq:alpahpm}) can be 
determined to a good accuracy by means of those 
entering $B\to \rho e\nu$ decays, leading 
to a precise determination of the point $s_0$
where $\cA^{(\bar B)}_{FB}(s_0)=0$ 
\cite{burdman0}. The interest in the zero of 
$\cA^{(\bar B)}_{FB}(s)$ 
is further reinforced by the fact that most 
of the intrinsic hadronic uncertainties
affecting $T_{1,2}$, $A_1$ and $V$ 
cancel in $\alpha_\pm(s)$ \cite{ABHH99,burdman0}, an
observation that can be justified in the large-energy expansion 
of heavy-to-light form factors  \cite{LEET}.
In this limit it is also easy to realize that 
$|\alpha_-(s)/\alpha_+(s)|=r_\Ks/(1-s) \ll 1$,
so that  the term proportional 
to $\ct^\prime$ in (\ref{eq:dfbabvllex})
is to a good approximation negligible. 
Since the position of $s_0$ 
does not depend on magnitude or 
sign of $C_{10}$ (assuming $C_{10}\not=0$)
we conclude that  within our 
New Physics scenario the zero of 
$\cA^{(\bar B)}_{FB}(s)$ remains 
unchanged with respect to the SM case
($s_0|_{\rm SM} = 0.10^{+0.02}_{-0.01}$ \cite{ABHH99}).

\begin{figure}[t]
\vskip 0.0truein
\centerline{\epsfysize=3.5in   
{\epsffile{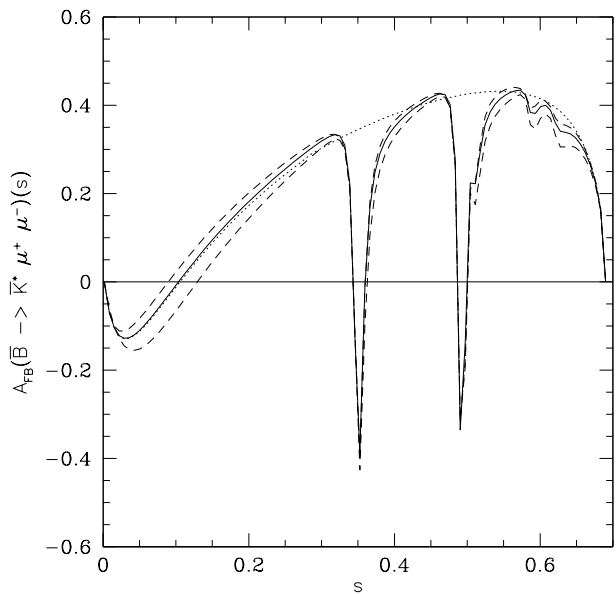}}}
\vskip 0.0truein
\caption[]{ \it Forward-backward asymmetry for
$\bar B\to \bar K^* \mu^+\mu^-$, defined as in
(\ref{eq:asdef}).
The solid (dotted) curves have been
obtained employing the Kr\"uger-Sehgal approach
(using  $\cne(s) \equiv C_9^{\rm EP}(s)$~). The
dashed lines show the effect of varying the
renormalization scale of the Wilson Coefficients
between $m_b/2$ and $2 m_b$,
within the Kr\"uger-Sehgal approach.}
\label{fig:AFB}
\end{figure}

Contrary to $s_0$, magnitude and sign of the forward-backward 
asymmetry can be very much affected by  possible 
non-standard contributions to $C_{10}$. 
The sign, in particular, is of great interest being 
related in a model-independent way to the relative signs 
of the Wilson coefficients. 
This relation deserves a clarifying discussion, as there
is apparently some confusion on this issue in the recent literature.
\begin{itemize}
\item{}
First of all we stress that the sign is different for 
$B$ and $\bar B$ decays. In fact, in the limit of 
$CP$ conservation one expects 
\beq
\cA^{(\bar B)}_{FB}(s) = - \cA^{(B)}_{FB}(s)~.
\eeq
This can be easily understood by noting that $CP$ 
conjugation requires not only the exchange $b \leftrightarrow
\bar b$ but also the one of $\mu^+ \leftrightarrow \mu^-$. 
Since the two leptons are emitted back to back 
in the dilepton center-of-mass frame, 
the asymmetry defined in terms of the direction of the 
positive charged lepton (both for $B$ and $\bar B$), 
changes sign under $CP$ conjugation.
\item{}
The sign in (\ref{eq:dfbabvllex}) implies that within the SM, 
where $\Re(C_{10}^*C_9) < 0$,
$\cA^{(\bar B)}_{FB}(s)$ is {\it positive\/} for $s > s_0$
(see Fig.~\ref{fig:AFB}).
This coincides with the SM behavior of the 
inclusive forward-backward asymmetry of the process
$b \to s \mu^+ \mu^-$ (see e.g. \cite{BI})
and indeed it has a simple partonic interpretation
(we recall that we denote by  $\bar B$ the meson with a valence $b$
quark).
At sufficiently large values of $q^2$ the contribution of $C_7$ 
can be neglected and, within the SM, the decay is almost 
a pure  $(V-A)\times(V-A)$ interaction 
($C_{10}|_{SM} \approx - C_9)$. In the $\bar B$
rest frame the emitted $s$ quark tends to be left-handed 
polarized and, when its spin 
is combined with the one of the spectator, this leads to a $\bar K^*$
meson with helicity $-1$ or $0$. Since the initial $\bar B$ 
meson has spin $0$, the total helicity of the 
recoiling lepton pair must also be $-1$ or $0$, respectively. 
If it is zero then there is no forward-backward 
asymmetry, as in the $\bar B \to \bar K \mu^+\mu^-$ 
case. On the other hand, if the polarization 
of the lepton pair is $-1$, then the positive lepton prefers
to travel backward with respect to the total 
momentum of the dilepton system, or in the direction of 
the $K^*$ meson.
This configuration corresponds to a positive $\cos\theta$, 
leading to a positive $\cA^{(\bar B)}_{FB}(s)$. 
\end{itemize}
Having firmly established the sign of  $\cA^{(\bar B)}_{FB}(s)$
within the SM, a striking signal of New Physics
could clearly be observed if ${\rm sgn}(\Re C_{10})= -
{\rm sgn}(\Re C_{10}|_{\rm SM})$.
In this case $\cA^{(\bar B)}_{FB}(s)$
would be positive for $s < s_0$ and 
negative for $s > s_0$, opposite to the  
SM expectation. Similarly, a clear signal of 
non-standard dynamics would occur if 
$\Re C_{10}$ was purely imaginary, so 
that $\cA^{(\bar B)}_{FB}(s)$ would be 
very much suppressed with respect to 
the SM case. Note that in both of these examples
one could still have an absolute 
value of $C_{10}$ close to its SM 
expectation, hiding these New Physics 
effects in branching ratios and 
dilepton spectra.

\subsubsection{Forward-backward $CP$ asymmetry}
More generally, a powerful tool to probe a 
possible $CP$-violating phase in
$C_{10}$ is provided by the sum
of the forward-backward asymmetries
of $\bar B$ and $B$ decays, which is
expected to vanish in the absence of 
$CP$ violation.\footnote{~This effect has already 
been pointed out in Ref.~\cite{FBCPKruger} in the 
context of $b\to d\ell^+\ell^-$ transitions.} 
For this purpose we 
introduce the {\em forward-backward $CP$ asymmetry}, 
defined as
\beq
 \cA^{CP}_{FB}(s) = \frac{ \cA^{(\bar B)}_{FB}(s) +
 \cA^{(B)}_{FB}(s)}{\cA^{(\bar B)}_{FB}(s) - \cA^{(B)}_{FB}(s)}~.  
 \label{eq:FBCPdef}
\eeq
This observable  is very small within the SM, where the $CP$-violating 
phases of the relevant Wilson coefficients are 
suppressed by the factor $\Im(V_{ub} V_{us}^*/V_{tb}V^*_{ts})
\sim \cO(\eta \lambda^2) \sim 0.01$. The 
explicit calculation of $\cA^{CP}_{FB}(s)$ within the SM 
requires to keep the small $u\bar u$ contribution 
in $\cne(s)$ (see e.g. \cite{AHEPJC}), which we have so far neglected.
Employing the partonic calculation for both  
$u\bar u$ and $c\bar c$ loops we find 
\begin{eqnarray}
 \left. \cA^{CP}_{FB}(s)\right|_{SM} &=&
 \frac{\Im (V_{ub} V_{us}^*) }{\Re (V_{cb}V^*_{cs})} 
 \frac{\Im \left[ h\left(\frac{m_c}{m_b},\frac{m^2_B}{m^2_b}s\right) 
                - h\left(0,\frac{m^2_B}{m^2_b}s\right) \right] 
  (3C_1+C_2)}{\Re \cne(s)}\nonumber \\
 &&\times\left[ 1 + \frac{\alpha_+(s)}{s} ~
 \frac{m_b C_7  }{ m_B \Re\cne(s) } \right]^{-1}~,
\label{eq:FBCPSM}
\end{eqnarray}
which in the region above the $\Psi'$ peak  
leads to an integrated asymmetry below $10^{-3}$.

\begin{figure}[t]
\vskip 0.0truein
\centerline{\epsfysize=3.5in   
{\epsffile{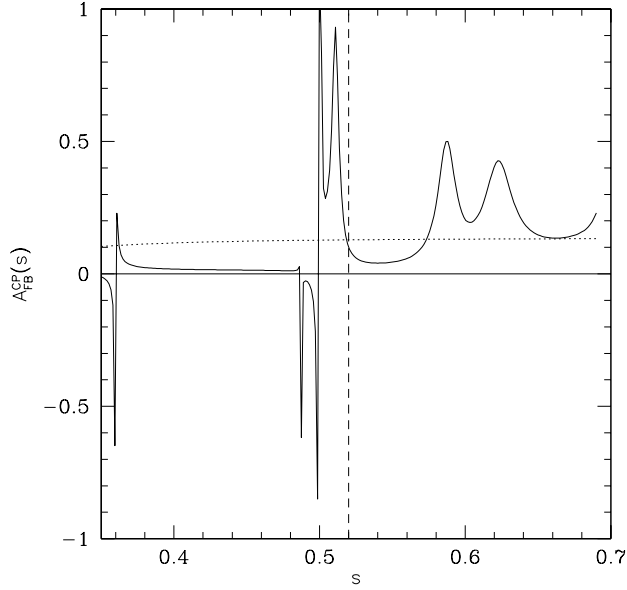}}}
\vskip 0.0truein
\caption[]{ \it The forward-backward 
$CP$ asymmetry defined in (\ref{eq:FBCPdef}{}), 
in units of $\Im \ct/\Re \ct$, as a function of $s$. 
Solid and dotted lines correspond to the 
Kr\"uger-Sehgal approach and to the choice
$\cne(s) \equiv C_9^{\rm EP}(s)$, respectively.
The vertical dashed line denotes the lower limit of the 
integration range in (\ref{eq:highq2}). }
\label{fig:fbcpkst}
\end{figure}

On the other hand,  if we allow $C_{10}$ to have a large 
$CP$-violating phase and neglect those of  
$C_7$ and $C_9$, as expected within our 
generic non-standard framework, we find 
\beq
 \cA^{CP}_{FB}(s) =   
 \frac{\Im \ct}{\Re \ct} \frac{\Im \cne(s)}{\Re \cne(s)}
 \left[ 1 + \frac{\alpha_+(s)}{s} ~
 \frac{m_b C_7  }{ m_B \Re\cne(s) } \right]^{-1}~,
\label{eq:FBCP}
\eeq
which can be substantially different from zero 
above the $c \bar c$ threshold if $\Im \ct/\Re \ct \sim \cO(1)$.
Note that the expression (\ref{eq:FBCP})  is almost free from 
uncertainties in the form factors, since for
large $s$ (where $\Im \cne(s)\not=0$)
the term proportional to $C_7$ is rather small. 
Unfortunately this virtue is somewhat compensated 
by the uncertainties in $\Im \cne(s)$ 
discussed in Section~\ref{sect:cne}.
A plot of $\cA^{CP}_{FB}(s)$, in units of $\Im \ct/\Re \ct$,
in the interesting region above the $\Psi$ peak
is shown in Fig.~\ref{fig:fbcpkst}.

To decrease the effect of the non-perturbative uncertainties in 
$\Im \cne(s)$ it
is convenient to integrate $\cA^{CP}_{FB}(s)$ over a large
interval of $q^2$. To avoid the uncontrollable errors associated with 
the narrow $\Psi$ and $\Psi^\prime$ peaks, as well as with the 
$D-\bar D$ threshold, we consider a safe 
integration region
\beq
\label{eq:highq2}
q^2_{\rm min} = 14.5~\mbox{GeV}^2 \leq q^2 < (m_B-m_{K^*})^2 = q^2_{\rm max}~,
\eeq
where we find 
\beq
 \Delta \cA^{CP}_{FB} =
 \int_{s_{\rm min}}^{s_{\rm max}}  ds
 \cA^{CP}_{FB}(s)  = (0.03 \pm 0.01) \times \frac{\Im \ct}{\Re \ct}~.
\label{eq:FBCPint}
\eeq
The central value in (\ref{eq:FBCPint}) has been obtained 
within the Kr\"uger-Sehgal approach, whereas the error has been 
estimated by comparing this result with the one obtained by
identifying $C_9^{\rm eff}(s)$ with $C_9^{\rm EP}(s)$.
Here and in Fig.~\ref{fig:fbcpkst} we did not use any
phenomenological correction factors for the resonance
contributions in applying the KS method,
that is we put $\kappa_V=1$ (notation of \cite{ks96}).

Unfortunately the numerical coefficient of $\Im \ct/\Re \ct$
in $\Delta \cA^{CP}_{FB}$ is rather suppressed, however 
it leaves open the possibility of $\cO(10\%)$ effects.
These would naturally occur if the non-standard 
contributions to  $Z^L_{bs}$ had the same magnitude 
as the SM term and a $CP$-violating phase of 
$\cO(1)$, a scenario that is allowed in  
most of the specific models discussed in Section~\ref{sect:3}.

\section{$B_s \to \mu^+ \mu^-$}
The constraint (\ref{eq:ctctpbound}) implies also 
an upper bound for $\Br(B_s \to \mu^+ \mu^-)$
in our generic non-standard scenario. Introducing the 
$B_s$ decay constant, $f_{B_{s}}$, the 
decay rate for this process can be written as 
\begin{eqnarray}
\Gamma(B_s \to \mu^+ \mu^-)&=&\frac{G_F^2 \alpha^2}{16 \pi^3}
f_{B_{s}}^2 |V_{ts}^\ast  V_{tb}|^2 m_{B_{s}} m_\mu^2 
 \left( 1- \frac{4 m^2_\mu}{ m^2_{B_{s}}}\right)^{1/2}
\left| \ct-\ct^\prime \right|^2~,
\end{eqnarray}
implying 
\begin{eqnarray}
\Br( B_s \to \mu^+ \mu^-) = \Br( B_s \to \mu^+ \mu^-)|_{\rm SM} \times
\left| \frac{\ct - \ct^\prime }{\ct\vert_{SM} } \right|^2~.
\label{eq:Brmm2}
\end{eqnarray}
Using the constraint (\ref{eq:ctctpbound}) we then find a maximal 
enhancement of a factor 7 for $\Br( B_s \to \mu^+ \mu^-)$
with respect to the SM value.

Employing the full next-to-leading order expression for 
$\ct|_{\rm SM}$ \cite{BBL,MU,BB99} one has
\begin{eqnarray}
\left. \Br( B_s \to \mu^+ \mu^-)\right|_{\rm SM} = 3.4 \times 10^{-9}
\left( \frac{f_{B_s}}{0.210 \mbox{GeV}} \right)^2
\left( \frac{|V_{ts}|}{0.040}           \right)^2
\left( \frac{\tau_{B_s}}{1.6 \mbox{ps}} \right)
\left( \frac{{\overline m}_t(m_t) }{170 \mbox{GeV} } \right)^{3.12}~.
\label{eq:BrmmSM}
\end{eqnarray}
Allowing for the maximal enhancement in (\ref{eq:Brmm2})
and adopting conservative upper bounds for the ratios 
in (\ref{eq:BrmmSM}) we finally obtain 
\beq
\Br( B_s \to \mu^+ \mu^-) < 3.4 \times 10^{-8}~, 
\eeq
which is about two orders of magnitude below the current best 
limit from CDF \cite{CDF98}: 
$\Br( B_s \to \mu^+ \mu^-) < 2.6 \times 10^{-6}$ (95\% C.L.).

\section{Summary and conclusions}

We have presented a study of the rare decay modes
$B\to K^{(*)}\nu\bar\nu$, $B\to K^{(*)}\ell^+\ell^-$ and $B_s\to\mu^+\mu^-$,
which are mediated by $b\to s$ FCNC transitions. These processes
have long been recognized as very interesting probes of the flavour
sector where New Physics effects could modify considerably the 
Standard Model expectations.

In this paper we have pursued the idea that the largest
deviations from the Standard Model could arise in the FCNC
couplings of the $Z$ boson. We have thus investigated a scenario
where new dynamics determines the 
$\bar s_{L,R}\gamma^\mu b_{L,R} Z_\mu$ interactions, while the
contributions of a different origin (boxes, photonic penguins)
are still, to a good approximation, given by their Standard Model
values.
As we have shown, this scenario is both phenomenologically 
and theoretically well motivated. Indeed, contrary to other 
FCNC amplitudes, 
the $\bar sbZ$ couplings are not yet very well
constrained by experimental data and considerable room for substantial
modifications still exists. On the other hand, 
also on theoretical grounds these couplings play a special role 
and are potentially dominant
in the presence of a high scale of New Physics.
It has also been shown that such a generic scenario 
could naturally arise in specific and consistent extensions 
of the SM, as for instance in the framework of Supersymmetry.

Within the Standard Model the following branching ratios
are expected, listed here in comparison with the current
experimental limits:
\begin{equation}
\ba{rclll}
\Br(B\to K\nu\bar\nu)&\approx& 4\times 10^{-6}\qquad 
 &(< 7.7\times 10^{-4} &\cite{delphi96}) \\
\Br(B\to K^*\nu\bar\nu)&\approx& 1.3\times 10^{-5}\qquad 
 &(< 7.7\times 10^{-4} &\cite{delphi96})\\
\Br(B\to K\mu^+\mu^-)^{n.r.}&\approx& 6\times 10^{-7} 
 &(< 5.2 \times 10^{-6}  &\cite{CDF})\\
\Br(B\to K^*\mu^+\mu^-)^{n.r.}&\approx& 2\times 10^{-6}
 &(< 4\times 10^{-6}   &\cite{CDF})\\
\Br(B_s\to\mu^+\mu^-)&\approx& 3\times 10^{-9} 
 &(< 2.6\times 10^{-6} &\cite{CDF98})
\ea
\end{equation}
The Standard Model estimates have at present hadronic
uncertainties of typically $\pm 30\%$.
Our generic New Physics scenario still allows for substantial
enhancements that could saturate the experimental bounds for
$B\to K^*\mu^+\mu^-$ and increase the remaining branching fractions
by factors of $5$ to $10$.

An observable of particular interest is the forward-backward
asymmetry ${\cal A}^{(\bar B)}_{FB}$ in $\bar B\to \bar K^*\mu^+\mu^-$
decay. This quantity is complementary to rate measurements and
can reveal non-standard flavourdynamics that might remain
invisible from the decay rates alone.
We have clarified the sign of the asymmetry within the Standard Model.
The sign (as a function of the dilepton mass) 
has the same behaviour in the exclusive channel
$\bar B\to \bar K^*\mu^+\mu^-$ as in the inclusive decay
$b\to s\mu^+\mu^-$. As we have shown, even for the hadronic
process $\bar B\to \bar K^*\mu^+\mu^-$ the sign of
${\cal A}^{(\bar B)}_{FB}$ can be fixed in a model-independent way.
This property provides us with an important Standard Model test.
The ``wrong'' sign of the experimentally
measured ${\cal A}^{(\bar B)}_{FB}$ would be a striking manifestation
of New Physics.
Such a test is comparable, and complementary, to determining the position
of the ${\cal A}_{FB}$ zero, whose usefulness as a clean probe
of New Physics has been stressed in the literature.
An interesting observation is that within our scenario of
non-standard $Z$ couplings the asymmetry ${\cal A}^{(\bar B)}_{FB}$ 
is likely to be affected, possibly including even a change of sign, 
while this class of New Physics would leave the 
${\cal A}^{(\bar B)}_{FB}$ zero essentially unchanged.

Finally, we have emphasized that the CP violating forward-backward
asymmetry ${\cal A}^{CP}_{FB}$ is an interesting probe of
non-standard CP violation in the $\bar sbZ$ couplings.
Potential effects are of order $10\%$, compared to an entirely
negligible Standard Model asymmetry of about $10^{-3}$. 

Similar observables can also be studied with inclusive
modes such as $b\to s\mu^+\mu^-$, which are theoretically
cleaner and could play an important role for precision tests
in the future. Nevertheless, on a shorter term the exclusive
channels are more accessible experimentally, in particular
at hadron machines. As we have seen,
exciting possibilities for tests of the flavour sector
exist also in this case in spite of, in general, 
larger hadronic uncertainties. 
The pursuit of these opportunities in rare $B$ decays
will certainly
contribute to a deeper understanding of flavour physics
in the Standard Model and beyond.

\section*{Acknowledgements}
We thank J. Hewett, S. Mele, M. Pl\"umacher, 
B. Richter and T. Rizzo for interesting discussions. 
The work of G.I. has been supported in part 
by the German Bundesministerium f{\"u}r 
Bildung und Forschung under contract 05HT9WOA0. 

\newpage

\end{document}